\documentclass[aps,twocolumn,preprintnumbers,superscriptaddress,floatfix,pra,10pt]{revtex4-2}
\usepackage{bbm}
\usepackage{bm}
\usepackage{amsmath}
\usepackage{amssymb}
\usepackage{empheq}
\usepackage{graphicx}
\usepackage{mathrsfs}
\usepackage{amsfonts}
\usepackage{amsthm}
\usepackage{color}
\usepackage{multirow}
\usepackage{bigints}
\usepackage{txfonts}
\usepackage{float}
\usepackage{url}
\usepackage{soul}
\usepackage{csquotes}
\usepackage[colorlinks,linkcolor=blue,anchorcolor=blue,citecolor=blue,urlcolor=blue]{hyperref}
\makeatletter \tolerance = 10000 \tolerance = 10000
\makeatother

\begin{document}
	
	\title{Generic skyrmion phase diagram in ferrimagnetic films}
	
	\author{M. V. Wijethunga}
	\affiliation{Physics Department, The Hong Kong University of Science and 
		Technology, Clear Water Bay, Kowloon, Hong Kong, China}
	
	\author{X. R. Wang}
	\email[Corresponding author: ]{phxwan@cuhk.edu.cn}
	\affiliation{School of Science and Engineering, Chinese University of 
		Hong Kong (Shenzhen), Shenzhen 51817, China}
	\date{\today}
	
\begin{abstract}

Ferrimagnetic skyrmions offer enhanced tunability due to antiferromagnetically coupled sublattices and reduced net magnetization.
In chiral magnetic films at zero magnetic field, skyrmion stability is commonly characterized by a dimensionless parameter $\kappa$,
yet its applicability to ferrimagnetic systems remains unclear, as most studies assume a fixed, strong inter-sublattice exchange
coupling $J$. Here we investigate how variations in $J$ govern relaxed stable and metastable ferrimagnetic skyrmion configurations
and introduce a dimensionless parameter $\zeta_{eff}$ to characterize the crossover between strong and weak inter-sublattice locking.
In the strong-coupling regime, inter-sublattice locking enables stabilization of skyrmion in a sublattice where 
intrinsic Dzyaloshinskii–Moriya interaction is absent while the other sublattice has finite DMI, yielding a sublattice with DMI-free ferrimagnetic skyrmions.
As $J$ decreases, this locking breaks down,
leading to independent sublattice behavior and the failure of an effective $\kappa$-based description. Our results establish a
unified framework linking inter-sublattice exchange and skyrmion phase stability in ferrimagnetic systems.
\end{abstract}

\maketitle
	
\section{Introduction}
\label{S1}

Skyrmions are topologically non-trivial spin textures characterized by the skyrmion number $Q = \iint \frac{1}{4\pi}\vec{m} \cdot (\partial_x \vec{m} \times \partial_y
\vec{m})\,dxdy$, where $\vec{m}$ is the unit vector of magnetization. They are stabilized by the Dzyaloshinskii-Moriya interaction (DMI) in systems without inversion symmetry
\cite{XS2018,R1,R2}, and have attracted considerable attention due to both fundamental interest and potential applications in spintronic devices \cite{R13, R14, Back2020, He2023,
Gnoli2021, Hrabec2017, Kiselev2011, Parkin2015, Zhang2015, Zhang2015b, Zazvorka2019, Luo2018, Zhang2015c, Behera2020, Zhu2018, Tomasello2014, Morshed2022}. Depending on the
interplay among exchange interaction, magnetic anisotropy, DMI, and external magnetic field, various (meta)stable spin textures can emerge, including isolated skyrmions,
helical stripes, skyrmion crystals (SkX), and polygonal skyrmions \cite{R3,R4,R5,R6,R7}. Among different material platforms, ferrimagnetic (FiM) systems, consisting of two
antiferromagnetically coupled magnetic sublattices with unequal moments, provide enhanced tunability for skyrmion stabilization and control. By varying composition or
temperature, both the net magnetization and angular momentum can be controlled and can vanish at the magnetization and angular momentum compensation points, respectively
\cite{R16,R17,R18,R29,R30}. Near these points, ferrimagnets exhibit dynamics similar to antiferromagnets, allowing faster magnetization response and efficient current-driven
motion \cite{R18,R29,R30,R31,R33}. In addition, the reduced net magnetization suppresses dipolar interactions, which improves skyrmion stability and enables smaller skyrmion
sizes \cite{R34,R35}. Experimental observations of skyrmions in FiM materials, including rare-earth–transition-metal alloys and multilayer systems, further highlight
their potential for applications due to low stray fields, high mobility, and room-temperature stability \cite{R18,R24,R29,R31,R37,R38,R_Woo2018,R_Legrand2020}.
Recent studies have also explored
FiM skyrmions in terms of their dynamics, stability, and behavior near compensation points using both experiments and two sublattice models
\cite{R18, R16, R_Woo2018, Barker2016, Back2020}, demonstrating their distinct advantages over ferromagnetic (FM) counterparts.

In ultra-thin chiral ferromagnets, the stability and morphology of skyrmions are well understood within a unified framework based on two dimensionless parameters,
$\kappa$ and $\kappa’$, which measure the relative strengths of curling and collinear interactions \cite{R2,R8,R9,R10,R11,R12}. It has been shown that these two parameters
fully determine the (meta)stable spin textures, independent of individual material parameters. A complete phase diagram constructed in the $\kappa\kappa’$-plane captures the
transition between isolated and condensed skyrmion phases, where $|\kappa’|=4$ separates the two regimes in the presence of a perpendicular magnetic field and in the absence
of magnetic anisotropy, while in the presence of magnetic anisotropy and in the absence of a magnetic field, $\kappa=1$ serves as the phase boundary between isolated skyrmions
with positive formation energy and condensed skyrmions with negative formation energy \cite{R8,R9,R10,R11,R12}. While such a framework provides a complete understanding for
FM systems, FiM skyrmions have not yet been described within an analogous unified description. Existing studies on FiM systems are typically
limited to specific material parameters or particular regimes, and do not provide a general phase diagram that captures the transition between isolated and condensed skyrmions
across the full parameter space. The presence of two coupled sublattices introduces additional degrees of freedom, making it unclear whether an effective description similar
to the $\kappa$–$\kappa’$ framework can be established. Establishing a generic phase diagram for FiM skyrmions is therefore essential for understanding their
stability and morphology and for guiding both theoretical and experimental investigations.

To address this, we extend the $\kappa-\kappa'$ framework to FiM chiral films by including the effects
of antiferromagnetic (AFM) inter-sublattice exchange and unequal sublattice parameters. Micromagnetic simulations are carried out using MuMax3 \cite{R27},
adapted for coupled two sublattice systems. In the strong inter-sublattice coupling regime, the resulting phase diagram
maps out the stable and metastable states
for FiM skyrmions in the absence of magnetic field and identifies the boundary separating isolated and condensed skyrmion phases. 
We further demonstrate that skyrmions can remain stable even when one sublattice is DMI-free, provided the other
sublattice retains finite DMI. This strong inter-sublattice exchange locks the two sublattices enabling a FiM skyrmion structure
and highlights a stabilization
mechanism that is absent in ferromagnets. In the weak inter-sublattice coupling regime, the phase diagram is governed by the independent
behavior of each sublattice, where the effective $\kappa$-based description
breaks down, and each sublattice independently relaxes into its own stable and metastable states, thereby preventing the stabilization of
skyrmions in a DMI-free sublattice.
This approach offers a theoretical foundation for understanding skyrmion morphology in ferrimagnets and provides guidance for
interpreting recent experimental results in multilayers and rare-earth–transition-metal systems \cite{R18,R_Woo2018,R_Legrand2020}.

The remainder of this paper is organized as follows. In Sec. \ref{S2}, we introduce the micromagnetic model for skyrmions in an ultra-thin FiM
film and derive the dimensionless formulation in terms of the effective parameters $a_\ell$, $d_\ell$, $\kappa_{\ell_{eff}}$, and $\zeta_{eff}$.
In Sec. \ref{S3},
we identify the skyrmion stability with $\zeta_{eff}$ and the separation between strong and weak inter-sublattice coupling regimes using the normalized order
parameter $\Delta_{rms}$. We first analyze the strong-coupling limit, where the two sublattices are rigidly locked and the system behaves as a FiM skyrmion.
In this regime, isolated and condensed skyrmion phases are separated solely
by $\kappa_{eff} = 1$ criterion, and the phase diagram is constructed in the $\kappa_{1_{eff}}\kappa_{2_{eff}}$-plane.
We then examine the weak-coupling regime, where the two sublattices relax independently. In this limit, the separation between isolated
and condensed skyrmion phases is governed solely by the intrinsic control parameter $\kappa_\ell=1$ of each sublattice, and the phase diagram is constructed
in the $\kappa_{1}\kappa_{2}$-plane. We further demonstrate
that under strong inter-sublattice coupling, skyrmions remain stable even when one sublattice has finite DMI while the other is DMI-free,
revealing a stabilization mechanism enabled by inter-sublattice locking.  Finally, we summarize the key results and their physical
implications in Sec. \ref{S4}.

\section{Model and methodology}
\label{S2}

We consider an ultra-thin chiral FiM film of thickness $d$ in the $xy$-plane, and in the absence of
an external magnetic field. 
The ferrimagnet consists of two antiferromagnetically coupled sublattices, whose spin textures are described by unit vectors
$\vec{m}_1$ and $\vec{m}_2$, with $|\vec{m}_1|=|\vec{m}_2|=1$, magnetizations of the two sublattices, respectively.
The total energy of the film in the absence of external magnetic field can be expressed as
\begin{equation}
\begin{split}
E=&\: \int \Bigg\{\sum_{\ell=1, 2} t_\ell\Big[A_\ell|\nabla \vec{m}_\ell|^2 +
D_\ell[(\vec{m}_\ell\cdot\nabla)m_{z_\ell}-m_{z_\ell}(\nabla\cdot\vec{m}_\ell)]+\\
&K_{u_\ell} (1- m^2_{z_\ell})-\frac{1}{2}\mu_0 M_{s_\ell} \vec{H}_{d_\ell}\cdot \vec{m}_\ell\Big]+dJ(\vec{m}_1\cdot\vec{m}_2)\Bigg\} 
\:dS,
\end{split}
\label{E1}
\end{equation}
where  $J>0$ is the AFM inter-sublattice spin coupling constant. $A_\ell$, $K_{u_\ell}$, $D_\ell$, $M_{s_\ell}$, and $t_\ell$ denote the
intra-sublattice FM exchange stiffness, perpendicular magnetocrystalline anisotropy,  interfacial
DMI strength, the saturation magnetization, thickness of sublattice $\ell$ ($\ell=1, 2$), respectively, and $\mu_0$ 
is the vacuum permeability. In contrast to the strong-coupling limit where
the inter-sublattice exchange only enforces antiparallel alignment, here we explicitly treat $J$ as a control parameter and
investigate how it affects the relaxed (meta)stable configurations obtained after numerical relaxation. We define the FiM state
with $\vec{m}_{1}=-\vec{m}_{2}=\pm \hat{z}$ as the zero energy state ($E=0$).

For an ultra-thin film whose thickness is much smaller than the exchange length, the demagnetization effect reduces to a local
contribution and can be absorbed into an effective perpendicular anisotropy \cite{XS2018}. In the thin film limit, 
the magnetostatic interaction is mainly caused by interfacial magnetic charges. As a result, it can be treated as a local
easy-plane contribution. In this description, the demagnetization energy depends on the out-of-plane components of the
magnetization in the same way as a uniaxial anisotropy term. Therefore, we incorporate the demagnetization contribution into
the anisotropy by using an effective anisotropy constant
of the form, $K_\ell = K_{u_\ell}  - \mu_0  M_{s_\ell}^2/2,$ where $\ell=1,2$.

By rearranging Eq. (\ref{E1}), the effective parameters naturally emerge as thickness-weighted averages,
$A_{eff} = (A_1t_1+A_2t_2)/d$ and $D_{eff} = (D_1t_1+D_2t_2)/d$, where $d=t_1+t_2$, consistent with previous work \cite{R_Luo2024}.
Following the same scaling procedure as in Refs. \cite{R8, R2}, we introduce the effective length scale $L_{eff}=(4A_{eff})/(\pi D_{eff})$.
After rescaling spatial coordinates as $x\to x/L_{eff}$ and $y\to y/L_{eff}$, the total energy can be written as
\begin{equation}
\begin{split}
E =&\: dA_{eff}\iint\Bigg\lbrace \sum_{\ell=1, 2} [a_\ell |\nabla \vec{m}_\ell|^2+ \frac{4}{\pi}d_\ell[(\vec{m}_\ell\cdot\nabla)m_{z_\ell}- 
\\
&m_{z_\ell}(\nabla\cdot\vec{m}_\ell)] +
\frac{1}{\kappa_{\ell_{eff}}}
(1- m^2_{z_\ell})] +\frac{2}{\zeta_{eff}}(\vec{m}_1\cdot\vec{m}_2)\Bigg\rbrace \:dxdy,
\end{split}
\label{E2}
\end{equation}
where $\kappa_{\ell_{eff}} = (\pi^2 D_{eff}^2)/(16A_{eff}K_{\ell_{eff}})$ with $K_{\ell_{eff}}=K_\ell t_\ell/d$, and $\zeta_{eff}=(\pi^2 D_{eff}^2)/(8A_{eff}J)$,
measures the relative strength of anisotropy and inter-sublattice exchange compared to chiral interactions,
respectively. In addition, the dimensionless ratios $a_\ell = (A_\ell t_\ell)/(A_{eff}d)$ and
$d_\ell = (D_\ell t_\ell)/(D_{eff}d)$ quantify the relative exchange stiffness and
DMI strength of each sublattice with respect
to the effective FiM values.
When $a_\ell = 1/2$, the exchange stiffness of sublattice $\ell$ contributes equally to the effective exchange stiffness of the ferrimagnet, whereas
$a_\ell > 1/2$ ($a_\ell < 1/2$) indicates that the sublattice contributes more (less) strongly to the total
exchange stiffness of the ferrimagnet. Similarly, $d_\ell = 1/2$ corresponds to a sublattice whose chiral interaction strength contributes equally
to the effective DMI strength, while $d_\ell > 1/2$ ($d_\ell < 1/2$) implies an enhanced (reduced) contribution to chiral twisting.
Differences in $a_\ell$ and $d_\ell$ therefore lead to unequal
characteristic length scales and stiffness within the two sublattices,
which can produce internal asymmetry in FiM skyrmion textures.
The (meta)stable configurations satisfy the following equation
\begin{equation}
\vec{m}_\ell\times\left\lbrace a_\ell\nabla^2\vec{m}_\ell+\frac{4}{\pi}d_\ell[(\nabla\cdot\vec{m}_\ell)\hat{z}-\nabla 
m_{z_\ell}]+\frac{1}{\kappa_{\ell_{eff}}}m_{z_\ell}\hat{z}-\frac{1}{\zeta_{eff}}\vec{m}_{\ell'}\right\rbrace=0,
\label{E3}
\end{equation}
here $\ell'=3-\ell$, and $x$ and $y$ are measured in the units of $L_{eff}$. 
From the aforementioned Eq. (\ref{E3}), it is evident that the stable and metastable skyrmion configurations are characterized by a set of
dimensionless parameters, $\kappa_{\ell_{eff}}$, $\zeta_{eff}$, and the relative exchange and DMI ratios,
$a_\ell$ and $d_\ell$, rather than by the individual micromagnetic parameters $A_\ell$, $D_\ell$, $M_{s_\ell}$, $K_{u_\ell}$, $J$, and the thickness $t_\ell$. This
dimensionless formulation provides a consistent framework for analyzing FiM skyrmion states and facilitates systematic comparisons between
different chiral magnetic materials in terms of effective parameters. While $\zeta_{eff}$ primarily quantifies the relative strength of inter-sublattice
coupling and controls the crossover between weakly and strongly coupled regimes, $\kappa_{\ell_{eff}}$ governs the
skyrmion phase stability,  while determining the precise
crossover position together with $a_\ell$ and $d_\ell$. 
This offers a precise mathematical framework for their evaluation and 
allows us to compare skyrmions in different chiral magnets in terms of effective parameters. 

The magnetization dynamics of the FiM system are governed by two coupled Landau-Lifshitz-Gilbert (LLG) equations for the
two sublattices ($\ell = 1,2$),
\begin{equation}
\frac{\partial\vec{m}_\ell}{\partial t} = -\gamma_{\ell}\left( \vec{m}_\ell\times\vec{H}_{eff,\:\ell}\right) +\alpha_{\ell} 
\left( \vec{m}_\ell\times\frac{\partial\vec{m}_\ell}{\partial t}\right) ,
\label{E4}
\end{equation}
where $\gamma_{\ell}$ and $\alpha_{\ell}$ are the gyromagnetic ratio and the Gilbert damping constant of sublattice $\ell$, respectively. The dynamics of the
two sublattices are coupled through the effective fields defined as $\vec{H}_{eff,\:\ell}=-1/(\mu_0 M_{s_\ell})\:\delta E/\delta \vec{m}_\ell$. The effective
magnetic field acting on sublattice $\ell$ is given by
\begin{equation}
\vec{H}_{eff,\:\ell}=\frac{2A_\ell}{\mu_0 M_{s_\ell}}\nabla^2
\vec{m}_\ell+\frac{2K_{u_\ell}}{\mu_0 M_{s_\ell}}m_{z_\ell}\hat{z}+\vec{H}_{d,\:\ell}+\vec{H}_{{DM},\:\ell}-\frac{J\vec{m}_{\ell'}}{\mu_0 M_{s_\ell}},
\label{E5}
\end{equation}
where $\ell' = 3 - \ell$. The first term is the exchange field, the second term is the crystalline magnetic anisotropy field, the
demagnetization field $\vec{H}_{d,\:\ell}$, the DMI field $\vec{H}_{DM,\:\ell}$, and the last term is the AFM inter-sublattice exchange field of
sublattice $\ell$. No external magnetic field is applied in this work. In the absence of energy sources, the dissipative nature of
the dynamics ensures that the total energy decreases monotonically with time \cite{R20,R21,R22}, allowing an initial configuration
to relax into a (meta)stable state.
The material parameters used in this work are chosen to be representative of experimentally studied FiM
thin films and multilayers. As examples of rare-earth transition-metal ferrimagnets, $\mathrm {DyCo_3}$-based films
have been reported with exchange stiffness on the order of a few $\mathrm{pJ\,m^{-1}}$, perpendicular anisotropy on
the order of $10^2\, \mathrm {kJ\,m^{-3}}$, and net saturation magnetization on the order of $10^2$-$10^3\, \mathrm
{kA\,m^{-1}}$, together with interfacial DMI strengths on the order of $1\, \mathrm {mJ\,m^{-2}}$ \cite{R35}.
As examples of amorphous GdCo or CoGd FiM films, typical values include exchange stiffness of several
$\mathrm {pJ\,m^{-1}}$, perpendicular anisotropy of several $10$-$10^2\, \mathrm {kJ\,m^{-3}}$, and a much smaller
net magnetization (often $10$-$10^2\, \mathrm {kA\,m^{-1}}$), with interfacial DMI strengths typically below
$1\, \mathrm {mJ\,m^{-2}}$ depending on the adjacent heavy-metal layers \cite{R24}. As an example of a
rare-earth-free ferrimagnet, $\mathrm {Mn_4N}$ films have been reported with low net magnetization and perpendicular
anisotropy \cite{R25}, while the effective interfacial DMI can vary significantly with interface engineering.
These experimentally motivated parameter sets are used as realistic FiM examples to examine how varying
$J$ influences the relaxed (meta)stable skyrmion configurations.

\section{Results}
\label{S3}

\subsection{\texorpdfstring{Separation of strong and weak inter-sublattice coupling regimes in zero magnetic field}
	{Separation of strong and weak inter-sublattice coupling regimes in zero magnetic field}}
\label{S3A*}
\begin{figure*}[t] 
	\centering
	\includegraphics[width=17.8cm]{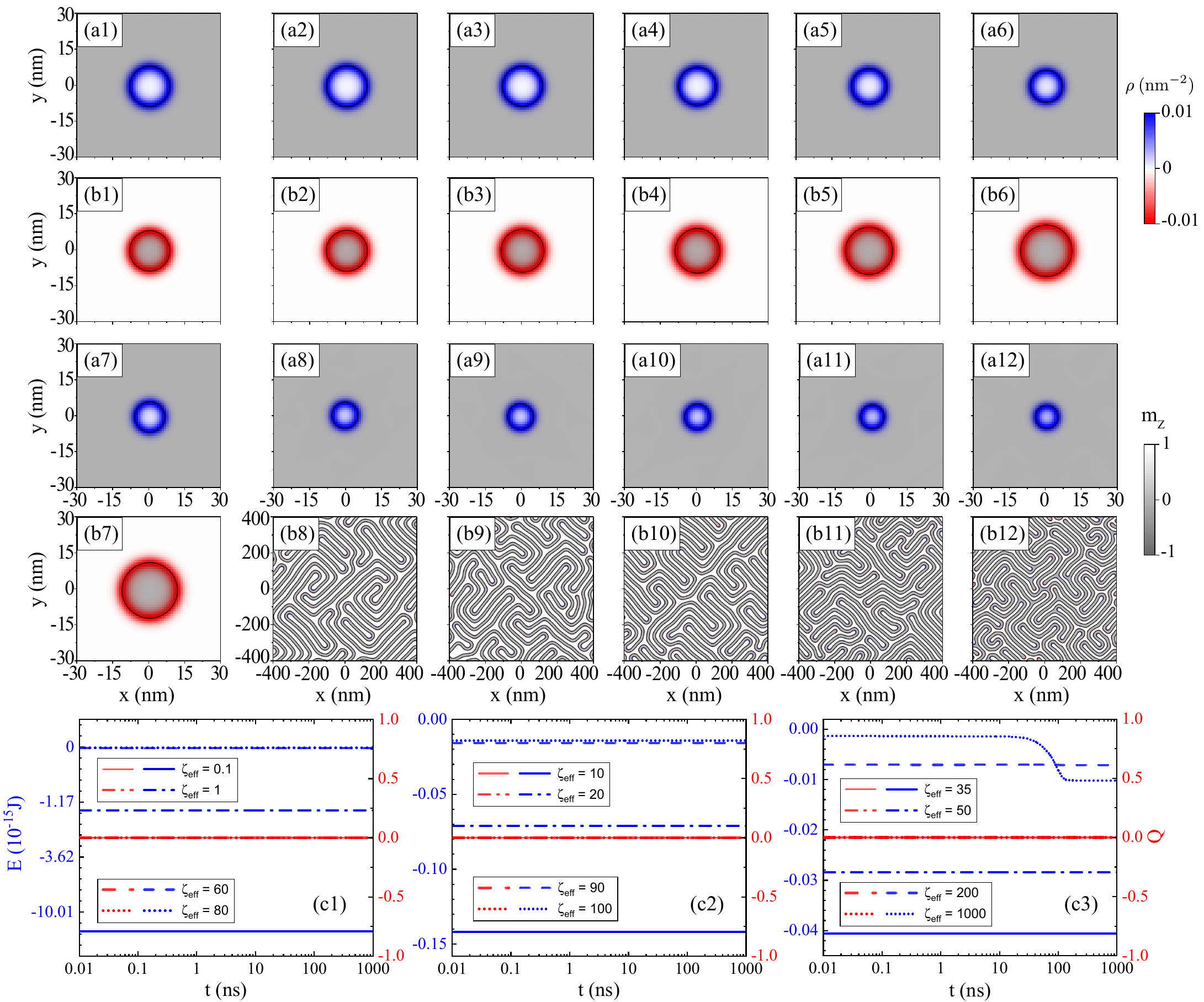} 
	\caption{(Meta)stable skyrmions for different values of $\zeta_{eff}$ for fixed $\kappa_{1_{eff}} = 1.4$ and $\kappa_{2_{eff}} = 2.6$. Skyrmions
			for sublattice 1 (\textbf{a1–a12}) and for sublattice 2 (\textbf{b1–b12}) in a sample of $800\,\mathrm{nm}\times800\,\mathrm{nm}\times2\,\mathrm{nm}$
			with saturation magnetizations $M_{s_1}= 0.58\, \mathrm{MA\,m^{-1}}$ and $M_{s_2}= 0.7\, \mathrm{MA\,m^{-1}}$, and exchange stiffness $A_\ell$ and DMI
			strength $D_\ell$ in Set 1 listed in Tab. \ref{T1} for $t_1 = t_2$. (\textbf{a1, b1}) $\zeta_{eff} = 0.1$. (\textbf{a2, b2}) $\zeta_{eff} = 1$.
			(\textbf{a3, b3}) $\zeta_{eff} = 10$.
			(\textbf{a4, b4}) $\zeta_{eff} = 20$. (\textbf{a5, b5}) $\zeta_{eff} = 35$. (\textbf{a6, b6}) $\zeta_{eff} = 50$. (\textbf{a7, b7}) $\zeta_{eff} = 60$.
			(\textbf{a8, b8}) $\zeta_{eff} = 80$. (\textbf{a9, b9}) $\zeta_{eff} = 90$. (\textbf{a10, b10}) $\zeta_{eff} = 100$. (\textbf{a11, b11}) $\zeta_{eff} = 200$.
			(\textbf{a12, b12}) $\zeta_{eff} = 1000$. (\textbf{a1–a7}) and (\textbf{b1–b7}) isolated skyrmions in the strong coupling regime. (\textbf{a8–a12}) isolated
			skyrmions and (\textbf{b8-b12}) stripe skyrmions in the weak coupling regime. The color bar denotes the skyrmion charge density $\rho$ for each sublattice,
			and the gray scale encodes $m_z$. (\textbf{c1–c3}) The net skyrmion number $Q$ (red lines) and energy $E$ (blue curves) vs time $t$ (in logarithmic scale)
			for various $\zeta_{eff}$ values. The net skyrmion number remains $Q = 0$.}
	\label{F10}
\end{figure*}
\begin{figure*}[t] 
	\centering
	\includegraphics[width=17.8cm]{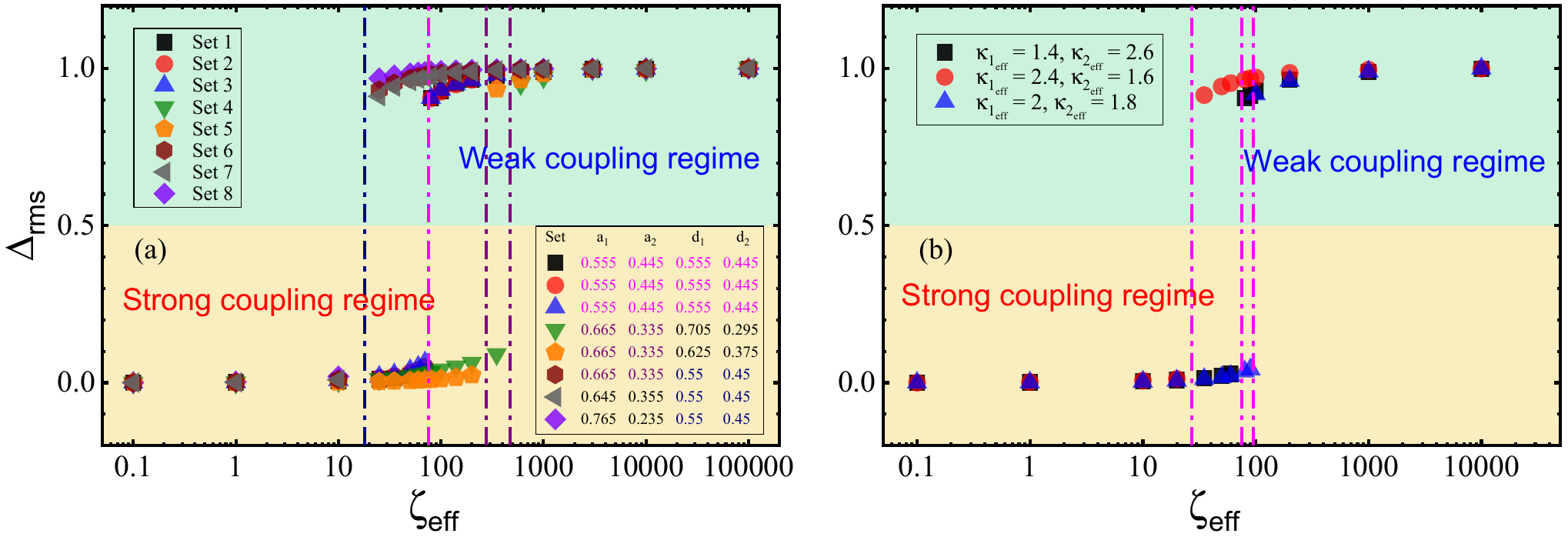} 
	\caption{Normalized order parameter, $\Delta_{rms}$ as a function of the effective inter-sublattice coupling parameter $\zeta_{eff}$ (in logarithmic
			scale) for $t_1 = t_2$. (a) For several representative material parameter sets ($A_\ell, D_\ell$) in Set 1 (rectangles),
			Set 2 (circles), Set 3 (up-triangles), Set 4 (down-triangles), Set 5 (pentagons), Set 6 (hexagonals), Set 7 (left-triangles), and Set 8 (diamonds) listed in
			Tab. \ref{T1} for fixed $\kappa_{1_{eff}} = 1.4$ and $\kappa_{2_{eff}} = 2.6$. (b) For various values of $\kappa_{1_{eff}} = 1.4$,
			$\kappa_{2_{eff}} = 2.6$ (squares), $\kappa_{1_{eff}} = 2.4$, $\kappa_{2_{eff}} = 1.6$ (circles), and $\kappa_{1_{eff}} = 2$, $\kappa_{2_{eff}} = 1.8$ (up-triangles) 
			for a fixed set of material parameters $(A_\ell, D_\ell)$ corresponding
			to Set 1 in Tab. \ref{T1}.
			The quantity $\Delta_{rms}$ serves as an order parameter for inter-sublattice locking: $\Delta_{rms} \approx 0$ indicates strong AFM locking and a FiM skyrmion, while
			$\Delta_{rms} \approx 1$ corresponds to the breakdown of locking and independent relaxation of the two sublattices.}
	\label{F0}
\end{figure*}
First, we investigate how the strength of inter-sublattice exchange coupling governs the separation between strong and
weak coupling regimes in FiM skyrmion systems in the absence of an external magnetic field. In the strong coupling regime,
the two antiferromagnetically coupled sublattices remain locked, forming FiM skyrmions that behave as a single effective
magnetic structure (discussed in Sec. \ref{S3A}). As the coupling strength is reduced, this locking progressively weakens, leading to a breakdown of collective
behavior and allowing the sublattices to relax independently according to their own material parameters (discussed in Sec. \ref{S3B}). To quantitatively
characterize this crossover, we introduce the normalized order parameter $\Delta_{rms}=\sqrt{\left\langle |\vec{m}_1+\vec{m}_2|^2\right\rangle}/\sqrt{2}$
which measures the normalized root-mean-square deviation from perfect
antiparallel alignment between the two sublattice magnetizations $\vec{m}_1$ and $\vec{m}_2$. From Eq. (\ref{E3}), it is clear that the stationary
magnetization configurations are governed exclusively by the dimensionless parameters $a_\ell$, $d_\ell$, $\kappa_{\ell_{eff}}$, and $\zeta_{eff}$.
Since $\Delta_{rms}$ is computed from the equilibrium solutions $\vec{m}_1$ and $\vec{m}_2$, it follows that $\Delta_{rms}
=f\,\!(a_\ell, d_\ell, \kappa_{\ell_{eff}}, \zeta_{eff})$, where the dependence arises implicitly through the solutions of Eq. (\ref{E3}).
With this normalization,
$\Delta_{rms}=0$ corresponds to perfectly antiparallel alignment, while $\Delta_{rms}=1$ indicates complete decoupling of the
two sublattices. In the strong-coupling limit, $\Delta_{rms}\approx0$, reflecting coherent locking of the sublattices.
As the inter-sublattice exchange coupling is reduced, the energetic cost of relative distortions becomes comparable to the
intrinsic exchange, DMI, and anisotropy energies within each sublattice, allowing the two sublattices to relax independently and ultimately leading to the weak-coupling limit,
$\Delta_{rms}\approx1$. This
crossover is marked by a rapid increase of $\Delta_{rms}$ beyond a critical value of $\zeta_{eff}$, signaling the breakdown of
inter-sublattice locking, and indicating that the transition between strong and weak coupling regimes is primarily governed by the effective coupling $\zeta_{eff}$. While
$\zeta_{eff}$ sets the dominant scale separating the two regimes in zero magnetic field, the critical $\zeta_{eff}$ at which the crossover occurs is shifted by the intrinsic
dimensionless parameters $a_\ell$, $d_\ell$, and $\kappa_{\ell_{eff}}$ of each sublattice.

To examine how (meta)stable skyrmion configurations change with $J$, we use MuMax3 to obtain the (meta)stable states for $A_\ell$, $D_\ell$ as specified in Set 1 of Tab. \ref{T1}
for fixed $\kappa_{1_{eff}}=1.4$ and $\kappa_{2_{eff}}=2.6$ obtained by varying $K_{u_1}$ and $K_{u_2}$ in a sample of $800\,\mathrm{nm}\times800\,\mathrm{nm}\times2\,\mathrm{nm}$
with $M_{s_1}= 0.58\,\mathrm{MA\,m^{-1}}$ and $M_{s_2}= 0.7\, \mathrm{MA\,m^{-1}}$. The initial skyrmion configurations (same as in Sec. \ref{S3A}) relax into (meta)stable states
for various $\zeta_{eff}$ values specified in the caption obtained from the corresponding $J$ for the case of $t_1=t_2$. The resulting (meta)stable configurations are shown in
Fig. \ref{F10}(a1–a12) for sublattice 1 and Fig. \ref{F10}(b1–b12) for sublattice 2. The color bar represents the skyrmion charge density $\rho$, and the gray scale encodes $m_z$.
As $\zeta_{eff}$ increases, $J$ weakens and becomes insufficient to enforce collective behavior, so the intrinsic parameters of each sublattice become relevant, modifying the
skyrmion morphology and changing its size without altering the phase within each regime, leading to the transition from the strong- to the weak-coupling regime. For small
$\zeta_{eff}$ (large $J$), the strong-coupling regime is shown in Fig. \ref{F10}(a1–a7) and Fig. \ref{F10}(b1–b7), where isolated skyrmions are (meta)stable. In this regime,
the system is described by an effective parameter $\kappa_{eff}$, and the isolated skyrmion phase follows from $\kappa_{eff} < 1$, as discussed in Sec. \ref{S3A}. For large
$\zeta_{eff}$ (small $J$), the weak-coupling regime is governed by the intrinsic parameters $\kappa_\ell$, where each sublattice independently supports isolated skyrmions for
$\kappa_\ell < 1$ and condensed skyrmions for $\kappa_\ell > 1$ as discussed in Sec. \ref{S3B}, and this behavior is shown in Fig. \ref{F10}(a8–a12) and Fig. \ref{F10}(b8–b12),
where isolated and stripe skyrmions are (meta)stable, respectively. Importantly, $\zeta_{eff}$ only determines the coupling regime and does not affect the phase boundary 
separating isolated and condensed skyrmions within each regime, which is determined solely by $\kappa_{\ell_{eff}}$. The time evolution of the net skyrmion number $Q$ (the red
lines and the right $y$-axis) and total energy $E$ (the blue curves and the left $y$-axis) as shown in Fig. \ref{F10}(c1–c3), confirms that $Q$ remains zero for $t>0.1$ ns,
while the total energy converges to a stable value.

To further verify the crossover between strong and weak inter-sublattice coupling regimes, we numerically obtained the (meta)stable states for several representative material
parameter sets $A_\ell$, $D_\ell$ as specified in Set \# of Tab. \ref{T1} for the case $t_1 = t_2$ using the same method in order to calculate the normalized order parameter $\Delta_{rms}$
as a function of the effective inter-sublattice coupling parameter $\zeta_{eff}$ (in logarithmic scale) as shown in Fig. \ref{F0}. Figure \ref{F0}(a) shows the
results for several representative material parameter sets specified in the caption for fixed $\kappa_{1_{eff}}=1.4$ and $\kappa_{2_{eff}}=2.6$. For small $\zeta_{eff}$, all data
sets collapse near $\Delta_{rms} \approx 0$, indicating rigid antiparallel
alignment and the formation of a FiM skyrmion in the strong-coupling regime. As $\zeta_{eff}$ increases, a sharp crossover occurs, beyond which $\Delta_{rms}$ rapidly approaches
unity, signaling the breakdown of inter-sublattice locking and independent relaxation of the two sublattices ($\Delta_{rms} \approx 1$).  Although the precise crossover
location varies moderately across parameter sets (vertical dashed dot lines), the transition consistently occurs within a narrow range of
$\zeta_{eff}$ for each set. For parameter sets with identical dimensionless ratios $a_\ell$ and $d_\ell$, the square, circle, and up-triangle symbols collapse
into a single crossover, confirming that $\zeta_{eff}$ captures the dominant AFM exchange contribution governing
inter-sublattice locking as shown in Fig. \ref{F0}(a).
In contrast, when $a_\ell$ and $d_\ell$ differ, the crossover shifts, with $a_\ell$ leading to only
a very narrow shift for the sets with identical $d_\ell$, as shown by the hexagonal, left-triangle, and diamond symbols, while $d_\ell$ leads to a more
pronounced shift for the sets with identical $a_\ell$, as shown by the down-triangle, pentagon, and hexagonal symbols in Fig. \ref{F0}(a), indicating that the
dominant contribution arises from $d_\ell$, which controls the chiral twisting of the spin texture, reflecting $a_\ell$ and $d_\ell$ mainly affect
the position of the crossover, without altering the underlying coupling mechanism that determines the transition. Figure \ref{F0}(b) shows the results for various values of
$\kappa_{1_{eff}} = 1.4$, $\kappa_{2_{eff}} = 2.6$ (squares), $\kappa_{1_{eff}} = 2.4$, $\kappa_{2_{eff}} = 1.6$ (circles), and $\kappa_{1_{eff}} = 2$, $\kappa_{2_{eff}} = 1.8$
(up-triangles) for a fixed material parameter set $A_\ell, D_\ell$ corresponding to Set 1 in Tab. \ref{T1}, and shows that the crossover position also depends on $\kappa_{\ell_{eff}}$ and
shift as $\kappa_{\ell_{eff}}$ varies. Overall, it is clear that $\zeta_{eff}$ primarily governs the crossover
between the strong and weak coupling regimes, and that $\kappa_{l_{eff}}$ governs the skyrmion phase stability within each coupling regime while determining the precise
crossover position together with $a_\ell$ and $d_\ell$.

\subsection{\texorpdfstring{Strong coupling regime and ferrimagnetic skyrmions}
	{Strong coupling regime and ferrimagnetic skyrmions}}
\label{S3A}
\begin{figure*}[t] 
	\centering
	\includegraphics[width=17.8cm]{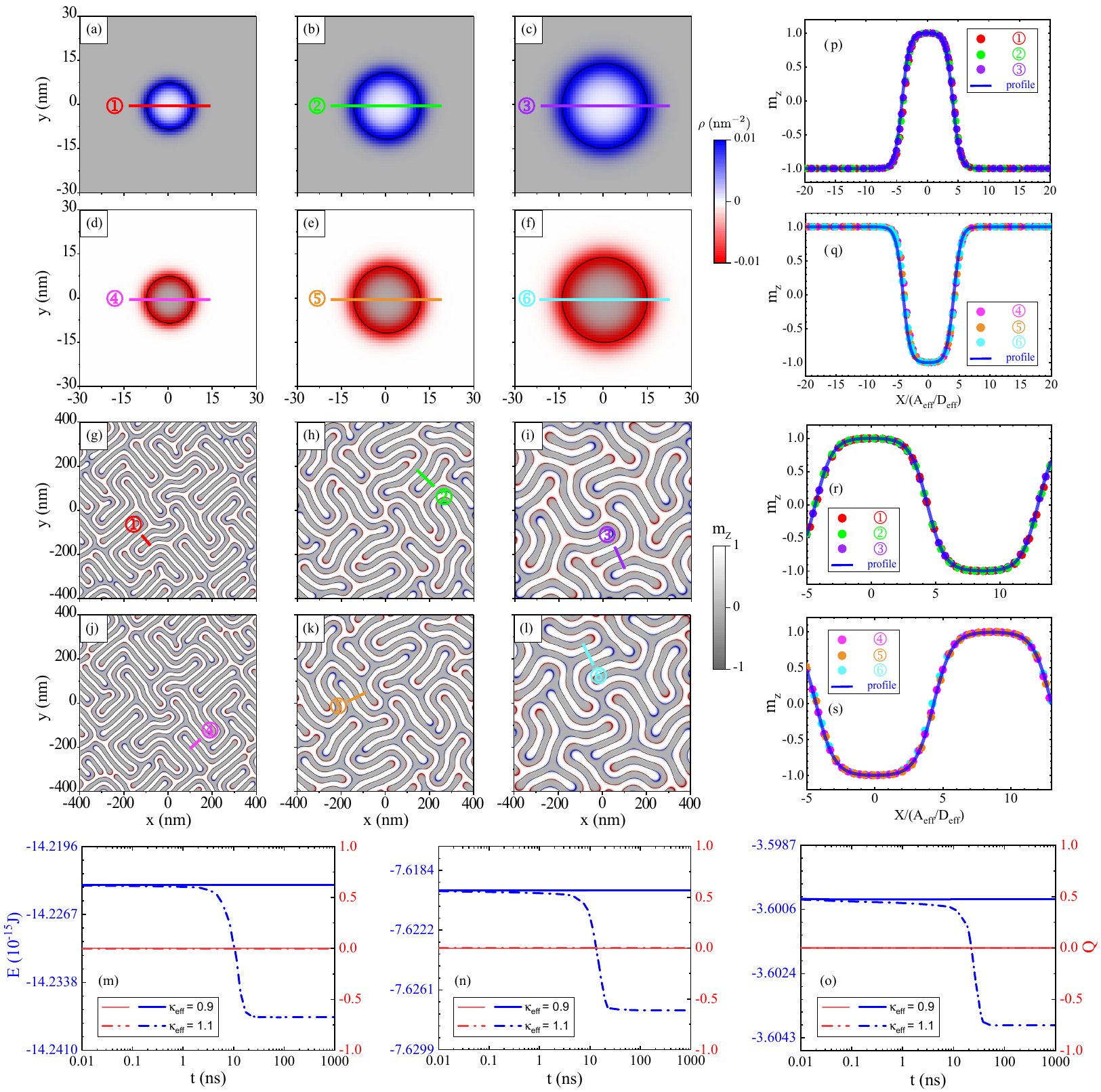} 
	\caption{(Meta)stable skyrmions in the strong coupling regime for different sublattice parameters $A_\ell, D_\ell$. Skyrmions for $\kappa_{eff}= 0.9$
		(\textbf{a-c}) for sublattice 1, (\textbf{d-f}) for sublattice 2 and 
		$\kappa_{eff}= 1.1$ (\textbf{g-i}) for sublattice 1, (\textbf{j-l}) for sublattice 2 in a sample of
		$800\,\mathrm{nm}\times800\,\mathrm{nm}\times2\,
		\mathrm{nm}$ with saturation magnetization $M_{s_1}= 0.58\, \mathrm{MA\,m^{-1}}$ and $M_{s_2}= 0.7\, \mathrm{MA\,m^{-1}}$, and 
		exchange stiffness $A_\ell$ and DMI strength $D_\ell$ in Set 1 (\textbf
		{a, d, g, j}), Set 2 (\textbf{b, e, h, k}), and Set 3 (\textbf{c, f, i, l}) listed in Tab. \ref{T1} for $t_1 = t_2$. 
		(\textbf{a-f}) Isolated skyrmions. (\textbf{g-l}) Stripe skyrmions.  The color bar denotes skyrmion 
		charge density $\rho$ for each sublattice, and the gray-scale encodes $m_z$.
		(\textbf{m-o}) The net skyrmion number $Q$ (red lines) and energy $E$ (in arcsinh scale, blue curves) 
		vs. time $t$ (in logarithmic scale) for $\kappa_{eff}= 0.9$ (the solid lines) and $\kappa_{eff}= 1.1$ (the dashed-dot lines) 
		with $A_\ell$ and $D_\ell$ in Set 1 (\textbf{m}),  Set 2 (\textbf{n}), and Set 3 (\textbf{o}). 
		(\textbf{p-s}) $m_z$-distribution along the red (the red symbols), green (the green symbols), purple (the purple symbols), 
		pink (the pink symbols), orange (the orange symbols), and blue (the blue symbols) 
		lines in (\textbf{a-c}), (\textbf{d-f}), (\textbf{g-i}), and (\textbf{j-l}), respectively. 
		The solid blue lines are $\Theta(r)=2\arctan[\sinh(r/w)/\sinh(R/w)]$ with $R=4.12$ and $w=1.17$ 
		(\textbf{p}), $R=4.12$ and $w=1.16$ (\textbf{q}), and $\Theta(x)=2\arctan[\sinh(|x|/w)/\sinh(L/2w)]$ with $L=8.67$ and $w=1.37$ 
		(\textbf{r}), $L=8.65$ and $w=1.35$ (\textbf{s}) where $\Theta$ is 
		the polar angle of magnetization, $R$ and $L$ are skyrmion size, and $w$
		is the skyrmion wall thickness. $r = 0$ and $x = 0$ denote skyrmion centers. The $x$-axis is in the units of $A_{eff}/D_{eff}$.}
	\label{F1}
\end{figure*}
We consider the strong inter-sublattice coupling regime, where the AFM exchange interaction
is sufficiently large ($J \gg 1$) to lock the two sublattices together during relaxation. In this limit, the inter-sublattice
exchange enforces a rigid antiparallel alignment between the two sublattices, such that both sublattices share
the same spatial magnetization profile. As a result, the FiM system behaves collectively, rather than
as two weakly interacting magnetic subsystems. Under strong coupling, the total energy reduces to the sum of the two sublattice energies evaluated
for a common spin texture. Consequently, the effective perpendicular anisotropy of the ferrimagnet is given by
$K_{eff}=K_{1_{eff}}+K_{2_{eff}}$.
By defining the effective control parameter that separate isolated skyrmion phase from condensed skyrmion phase in this regime as
$\kappa_{eff} = (\pi^2 D_{eff}^2)/(16A_{eff}K_{eff})$, one can have,
\begin{equation}
	\kappa_{eff}=\frac{1}{\left( \frac{1}{\kappa_{1_{eff}}}+\frac{1}{\kappa_{2_{eff}}} \right)}.
	\label{E6}
\end{equation}
Under the rigid antiparallel alignment, magnetization of sublattices 1, 2 can be written as $\vec{m}_1\approx-\vec{m}_2=\vec{m}$ ($m_{z_1}\approx-m_{z_2}= m_z$).
Therefore, the inter-sublattice locking term governed by $\zeta_{eff}$ vanishes identically due to the constraint $\vec m\times\vec m=0$. As a result,
Eq. (\ref{E3}) reduces to the two conditions $\vec m\times \vec{H}_{eff_1}=0$ and $\vec m\times \vec{H}_{eff_2}=0$,
where $\vec{H}_{eff_1}=a_1\nabla^2\vec{m}+\frac{4}{\pi}d_1[(\nabla\cdot\vec{m})\hat{z}-\nabla m_{z}]+\frac{1}{\kappa_{1_{eff}}}m_{z}\hat{z}$, 
and $\vec{H}_{eff_2}=a_2\nabla^2\vec{m}+\frac{4}{\pi}d_2[(\nabla\cdot\vec{m})\hat{z}-\nabla m_{z}]+\frac{1}{\kappa_{2_{eff}}}m_{z}\hat{z}$ 
that denote the sublattice-specific effective fields.
Since both conditions must be satisfied simultaneously, taking the sum of the two conditions yields a single
effective equation for the FiM skyrmion structure, $\vec m\times (\vec{H}_{eff_1}+\vec{H}_{eff_2})=0$. By considering the
identities $a_1+a_2=1$, $d_1+d_2=1$, and
Eq. (\ref{E6}), the (meta)stable configurations satisfy the following equation
\begin{equation}
	\vec{m}\times\left\lbrace \nabla^2\vec{m}+\frac{4}{\pi}[(\nabla\cdot\vec{m})\hat{z}-\nabla 
	m_{z}]+\frac{1}{\kappa_{eff}}m_{z}\hat{z}\right\rbrace=0.
	\label{E8}
\end{equation}
Equation (\ref{E8}) makes it clear that, under strong $J$, the phase stability is governed solely by $\kappa_{eff}$ for
each sublattice.

In FM systems in the absence of external magnetic field, $\kappa = 1$ marks the phase boundary between isolated and condensed skyrmion phases
\cite{R10,R11,R12,R13,R14} by corresponding to a vanishing domain wall energy. Under strong inter-sublattice coupling,
the FiM system behaves as a FiM skyrmion described by effective parameters, so the same
physical balance governs skyrmion stability and the critical condition remains $\kappa_{eff}= 1$.
\begin{figure*}[t] 
	\centering
	\includegraphics[width=17.8cm]{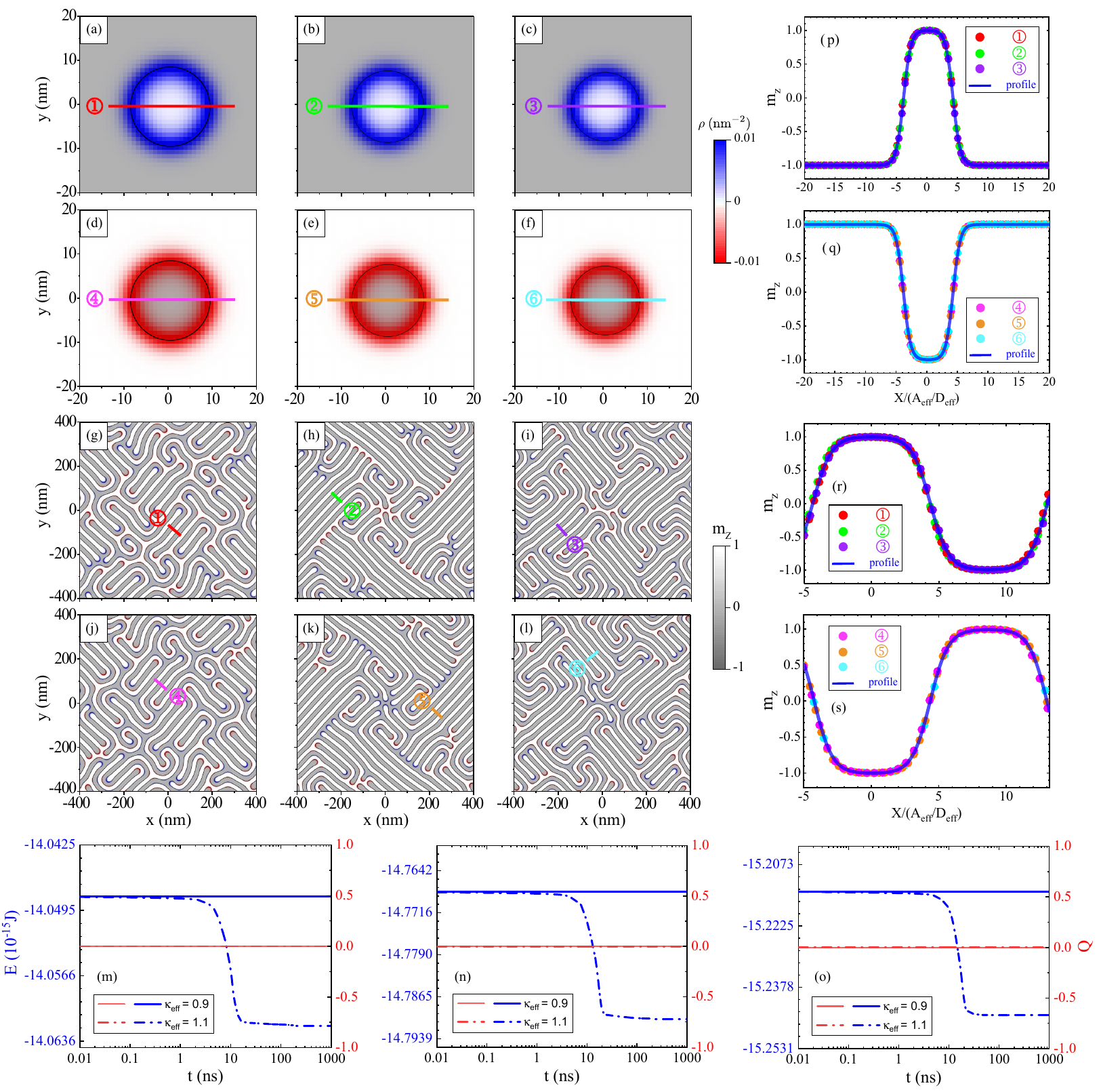} 
	\caption{(Meta)stable skyrmions in the strong coupling regime for different sublattice thickness $t_\ell$. Skyrmions for $\kappa_{eff}= 0.9$ (\textbf{a-c}) for sublattice 1,
			(\textbf{d-f}) for sublattice 2 and 
			$\kappa_{eff}= 1.1$ (\textbf{g-i}) for sublattice 1, (\textbf{j-l}) for sublattice 2 in a sample of
			$800\,\mathrm{nm}\times800\,\mathrm{nm}\times d$ with saturation magnetization $M_{s_1}= 0.58\, \mathrm{MA\,m^{-1}}$ and $M_{s_2}= 0.7\, \mathrm{MA\,m^{-1}}$, and 
			exchange stiffness $A_\ell$ and DMI strength $D_\ell$ in Set 6 listed in Tab. \ref{T1} for $t_1,t_2=1,1$ nm (\textbf
			{a, d, g, j}), $t_1,t_2=1,2$ nm (\textbf{b, e, h, k}), and $t_1,t_2=1,3$ nm (\textbf{c, f, i, l}). 
			(\textbf{a-f}) Isolated skyrmions. (\textbf{g-l}) Stripe skyrmions.  The color bar denotes skyrmion 
			charge density $\rho$ for each sublattice, and the gray-scale encodes $m_z$.
			(\textbf{m-o}) The net skyrmion number $Q$ (red lines) and energy $E$ (in arcsinh scale, blue curves) 
			vs. time $t$ (in logarithmic scale) for $\kappa_{eff}= 0.9$ (the solid lines) and $\kappa_{eff}= 1.1$ (the dashed-dot lines) 
			with $t_1,t_2=1,1$ nm (\textbf{m}),  $t_1,t_2=1,2$ nm (\textbf{n}), and $t_1,t_2=1,3$ nm (\textbf{o}). 
			(\textbf{p-s}) $m_z$-distribution along the red (the red symbols), green (the green symbols), purple (the purple symbols), 
			pink (the pink symbols), orange (the orange symbols), and blue (the blue symbols) 
			lines in (\textbf{a-c}), (\textbf{d-f}), (\textbf{g-i}), and (\textbf{j-l}), respectively. 
			The solid blue lines are $\Theta(r)=2\arctan[\sinh(r/w)/\sinh(R/w)]$ with $R=4.09$ and $w=1.2$ 
			(\textbf{p}), $R=4.05$ and $w=1.2$ (\textbf{q}), and $\Theta(x)=2\arctan[\sinh(|x|/w)/\sinh(L/2w)]$ with $L=8.61$ and $w=1.31$
			(\textbf{r}), $L=8.69$ and $w=1.37$ (\textbf{s}) where $\Theta$ is 
			the polar angle of magnetization, $R$ and $L$ are skyrmion size, and $w$
			is the skyrmion wall thickness. $r = 0$ and $x = 0$ denote skyrmion centers. The $x$-axis is in the units of $A_{eff}/D_{eff}$.}
	\label{F8}
\end{figure*}
\begin{figure*}[t] 
	\centering
	\includegraphics[width=17.8cm]{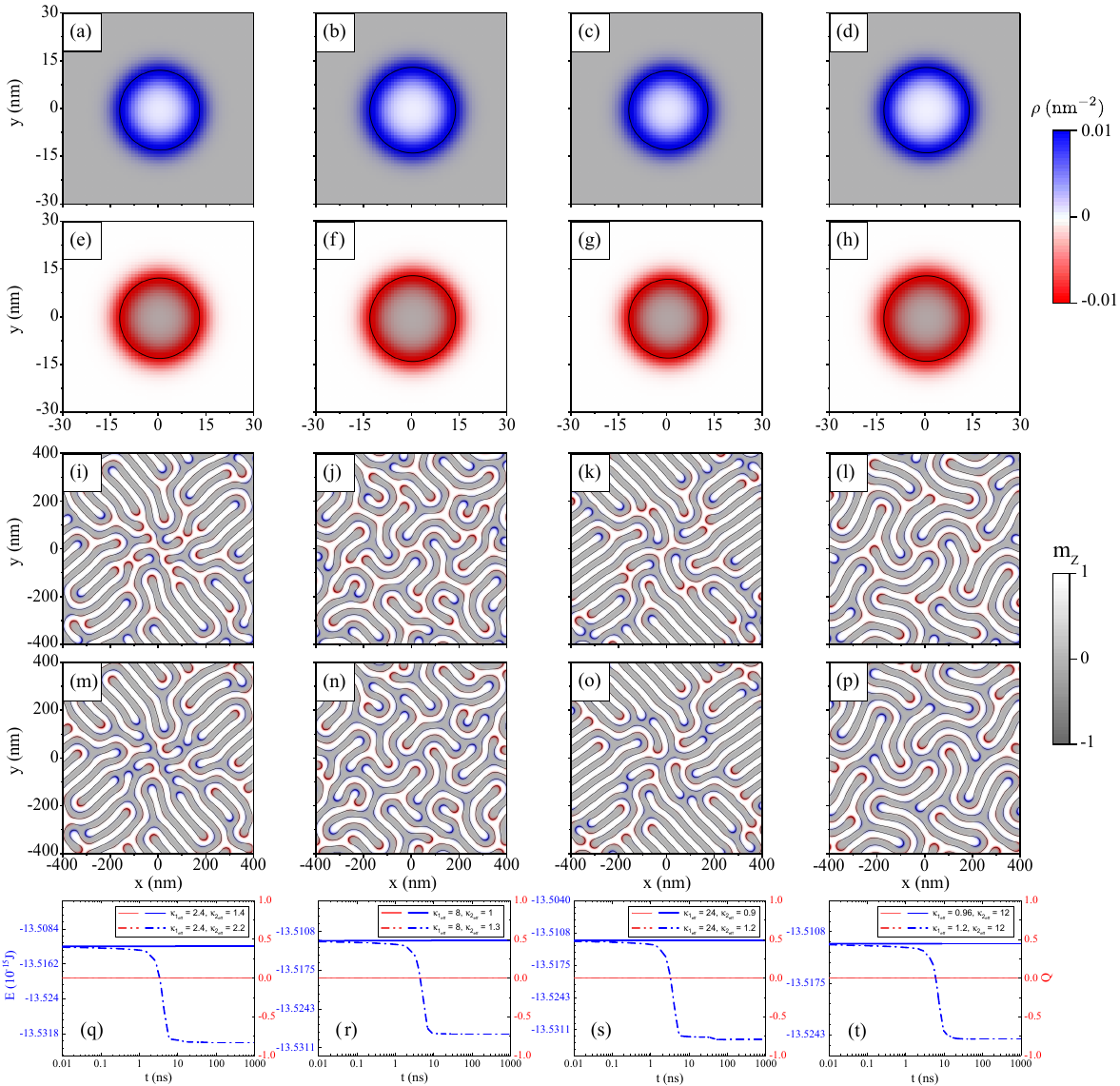} 
	\caption{(Meta)stable skyrmions for various $\kappa_{1_{eff}}$ and $\kappa_{2_{eff}}$ under strong coupling. Skyrmions in a sample of 
		$800\,\mathrm{nm}\times800\,\mathrm{nm}\times2\,\mathrm{nm}$ with exchange stiffness $A_\ell$ and 
		DMI strength $D_\ell$ in Set 9 in 
		Tab. \ref{T1} for $t_1=t_2$, and saturation magnetization $M_{s_1}= 0.58\, \mathrm{MA\,m^{-1}}$ and $M_{s_2}= 0.7\, \mathrm{MA\,m^{-1}}$. 
		(\textbf{a, e}) 
			$\kappa_{1_{eff}}=2.4$ and $\kappa_{2_{eff}}=1.4$.
			(\textbf{b, f}) $\kappa_{1_{eff}}=8$ and $\kappa_{2_{eff}}=1$. (\textbf{c, g}) $\kappa_{1_{eff}}=24$ and $\kappa_{2_{eff}}=0.9$. (\textbf{d, h})
			$\kappa_{1_{eff}}=0.96$ and $\kappa_{2_{eff}}=12$.
			(\textbf{i, m}) $\kappa_{1_{eff}}=2.4$ and $\kappa_{2_{eff}}=2.2$. 
			(\textbf{j, n}) $\kappa_{1_{eff}}=8$ and $\kappa_{2_{eff}}=1.3$.  
			(\textbf{k, o}) $\kappa_{1_{eff}}=24$ and $\kappa_{2_{eff}}=1.2$. (\textbf{l, p}) 
			$\kappa_{1_{eff}}=1.2$ and $\kappa_{2_{eff}}=12$. (\textbf{a-h}) Isolated skyrmions are (meta)stable. 
		(\textbf{i-p}) Stripe skyrmions are (meta)stable. The color bar denotes skyrmion charge density 
		$\rho$, and the gray-scale encodes $m_z$.
		(\textbf{q-t}) The net skyrmion number $Q$ (the red lines and the right $y$-axis) and energy $E$ in arcsinh 
		scale (the blue curves and the left $y$-axis) as a function of time (in logarithmic scale) for various 
		$\kappa_{1_{eff}}$ and $\kappa_{2_{eff}}$. Net skyrmion number is constant at $Q = 0$.}
	\label{F2}
\end{figure*}

To verify the critical value of $\kappa_{eff}=1$, we use MuMax3 to numerically
find critical $\kappa_{eff}$ in the absence of an external magnetic field for the case of $t_1 = t_2$.
The system is initialized with N\'{e}el-type skyrmion configurations in both sublattices, with antiparallel magnetization
orientations consistent with strong inter-sublattice AFM coupling and relaxed using the LLG dynamics. Unless
otherwise stated, the sample size is $800\,\mathrm{nm}\times800\,\mathrm{nm}\times2\,\mathrm{nm}$, with periodic boundary
conditions applied in the film plane. Following the spin dynamics given by the effective LLG equation, the initial
skyrmion configurations will eventually relax into the (meta)stable states for a given $K_{u_1}, K_{u_2}$, and $M_{s_1}= 0.58\,
\mathrm{MA\,m^{-1}}$, $M_{s_2}= 0.7\, \mathrm{MA\,m^{-1}}$, and $A_\ell$ and 
$D_\ell$ as specified in Set \# of Tab. \ref{T1}. A fixed $\zeta_{eff}= 0.1$ is chosen to ensure strong inter-sublattice
coupling by choosing the corresponding $J$,
and the data are collected by fixing $\kappa_{eff} = 0.9$ and $\kappa_{eff}= 1.1$, which are below and above the critical
value separating isolated and condensed skyrmion phases. The resulting (meta)stable
spin structures obtained from the simulations are shown in Fig. \ref{F1}(a, d, g, j) for Set 1, Fig. \ref{F1}(b, e, h, k) for Set 2,
and Fig. \ref{F1}(c, f, i, l) for Set 3, where Fig. \ref{F1}(a-c), and \ref{F1}(g-i) show the (meta)stable 
spin structures of sublattice 1, while Fig. \ref{F1}(d-f), and \ref{F1}(j-l) show the spin structures of
sublattice 2.
For $\kappa_{eff}= 0.9$, Fig. \ref{F1}(a-f) show isolated circular skyrmions, whereas for $\kappa_{eff} = 1.1$, Fig. \ref{F1}(g-l)
show stripe skyrmions. These results follow directly the critical value $\kappa_{eff} = 1$, which separates
isolated skyrmions from condensed skyrmions in the absence of an external magnetic field.
The skyrmion charge density 
$\rho=\frac{1}{4\pi}\vec{m_\ell}\cdot(\partial_x\vec{m_\ell}\times\partial_y\vec{m_\ell})$ is encoded by the colors specified in the color bar
for each sublattice. Owing to the antiparallel alignment of the two sublattices, the charge density carries opposite signs in the two
sublattices, while the skyrmion morphology remains identical under strong inter-sublattice coupling. As a consequence, the topological
charges associated with the two sublattices cancel each other, leading to a vanishing net skyrmion number $Q=0$.
Despite the distinct morphologies shown in Fig. \ref{F1}(a–l), all structures therefore have net skyrmion number
$Q=0$. For a specific $\kappa_{eff}$, the size of
the (meta)stable skyrmions varies with $A_{eff}$ and $D_{eff}$, or with the effective length scale $L_{eff}=(4A_{eff})/(\pi D_{eff})$.
To further validate this effective length scale, the $m_z$-distribution is examined along the red (red symbols), green (green symbols),
and purple (purple symbols) lines in Fig. \ref{F1}(a–c) for sublattice 1 and along the pink (pink symbols), orange (orange symbols),
blue (blue symbols)  lines in Fig. \ref{F1}(d–f) for sublattice 2, are plotted as shown in Fig. \ref{F1}(p), \ref{F1}(q), respectively.
Similarly, the $m_z$-distribution along the same sets of lines in Fig. \ref{F1}(g–i) for sublattice 1 and Fig. \ref{F1}(j–l) for
sublattice 2 are plotted as shown in Fig. \ref{F1}(r), \ref{F1}(s), respectively.
 The symbols are simulation data obtained from
MuMax3 and the $x$-axis is scaled by $A_{eff}/D_{eff}$. The spin profiles are well described by $\Theta(r)=2\arctan[\sinh(r/w)/\sinh(R/w)]$
for isolated circular
skyrmions \cite{XS2018} and $\Theta(x)=2\arctan[\sinh(|x|/w)/\sinh(L/2w)]$ for stripe skyrmions \cite{R10} as shown by the blue curves in
Fig. \ref{F1}(p), \ref{F1}(q) and \ref{F1}(r), \ref{F1}(s), respectively. Where $r = 0$ and $x = 0$ denote the skyrmion centers, $R$ and $L$ represent the skyrmion size
for circular and stripe cases, respectively, and $w$ is the domain wall thickness. The excellent agreement between simulations (symbols)
and theoretical profile (curves) confirms both the effective length scale proposed in this work and the universality of the skyrmion spin
profiles found in our earlier works \cite{R5, R10}. Moreover, the scaled isolated circular skyrmion sizes of sublattices 1 and 2 ($R = 4.12, 4.12$)
and the stripe skyrmion sizes ($L = 8.67, 8.65$) indicate that the skyrmions in both sublattices share the same characteristic size.
The (meta)stability and skyrmion nature of
these textures are further confirmed by the
relaxation dynamics shown in Fig. \ref{F1}(m-o), where the net skyrmion charge $Q$ (the red lines and the right $y$-axis) and total
energy $E$ (the blue curves and the left $y$-axis) are plotted as functions of time for $\kappa_{eff} = 0.9$  (the solid curves)
and $\kappa_{eff} = 1.1$ (the dashed-dot curves). In all cases, the net skyrmion number remains constant at $Q = 0$ for $t>10$ ps as each
sublattice carries a nonzero topological charge of opposite sign under strong antiparallel locking, while the total energy converges to
a stable value.

To further verify the critical value of $\kappa_{eff}=1$ for different sublattice thicknesses $t_\ell$, we use the same method as described above,
with the only changes being the sample size, taken as $800\,\mathrm{nm}\times800\,\mathrm{nm}\times d$ with $M_{s_1}= 0.58\,
\mathrm{MA\,m^{-1}}$, $M_{s_2}= 0.7\, \mathrm{MA\,m^{-1}}$, and the parameters $A_\ell$ and $D_\ell$ in
Set 6 of Tab. \ref{T1}, to numerically determine the critical $\kappa_{eff}$ in the absence of an external magnetic field for different $t_\ell$. The resulting (meta)stable
spin structures are shown in Fig. \ref{F8}(a, d, g, j) for $(t_1, t_2)=(1,1)$ nm, Fig. \ref{F8}(b, e, h, k) for $(t_1, t_2)=(1,2)$ nm, and 
Fig. \ref{F8}(c, f, i, l) for $(t_1, t_2)=(1,3)$ nm. Figure \ref{F8}(a–c) and \ref{F8}(g–i) correspond to sublattice 1, while Fig. \ref{F8}(d–f) and
\ref{F8}(j–l) correspond to sublattice 2. For $\kappa_{eff}=0.9$, Fig. \ref{F8}(a–f) show isolated circular skyrmions, whereas for $\kappa_{eff}=1.1$, Fig. \ref{F8}(g–l)
show stripe skyrmions. This directly confirms that $\kappa_{eff}=1$ separates isolated and condensed skyrmion phases even for different sublattice thicknesses in the
absence of an external magnetic field. The $m_z$-distribution is examined along the red (red symbols), green (green symbols),
and purple (purple symbols) lines in Fig. \ref{F8}(a–c) for sublattice 1 and along the pink (pink symbols), orange (orange symbols), blue (blue symbols)  lines in
Fig. \ref{F8}(d–f) for sublattice 2,
are plotted as shown in Fig. \ref{F8}(p), \ref{F8}(q), respectively. The spin profiles agree well with the known analytical forms for isolated circular and stripe
skyrmions \cite{XS2018, R10}, confirming both the effective length scale and the universality of the skyrmion profiles \cite{R5,R10}. Moreover, the scaled isolated circular
skyrmion sizes of sublattices 1 and 2 ($R = 4.09, 4.05$) and the stripe skyrmion sizes ($L = 8.61, 8.69$) indicate that the skyrmions in both sublattices share the same
characteristic size. The (meta)stability of these textures is confirmed by the relaxation dynamics in Fig. \ref{F8}(m–o), where the net skyrmion number remains zero and the
total energy converges to a stable value. This further shows that the sublattice thickness is properly incorporated into the effective description and that $\kappa_{eff}=1$
remains the phase boundary separating isolated and condensed skyrmion phases.
\begin{table}[htbp]
	\centering
	\setlength{\tabcolsep}{2mm}{
		\caption{14 sets of exchange stiffness $A_\ell$ and DMI strength $D_\ell$ of sublattices $\ell=1, 2$}
		\label{T1}	
		\begin{tabular}{lllll}
			\hline\hline\noalign{\smallskip}
			Set &$A_1\,(\,\mathrm{pJ\,m^{-1}}\,)$ &$A_2\,(\,\mathrm{pJ\,m^{-1}}\,)$ &$D_1\,(\,\mathrm{mJ\,m^{-2}}\,)$  & $D_2\,(\,\mathrm{mJ\,m^{-2}}\,)$ 
			\\ \noalign{\smallskip}\hline\noalign{\smallskip}
			1  & 4  & 3.2 & 2 & 1.6     \\ 
			2  & 4.2  & 3.36 & 1.5 & 1.2    \\	
			3  & 3.2  & 2.56 & 0.9 & 0.72     \\ 
			4  & 6 & 3 & 2.6 & 1.1   \\
			5  & 5.47 & 2.74 & 4.5 & 2.7    \\ 
			6  & 6 & 3 & 2.2 & 1.8   \\	 
			7  & 4.5 & 2.5 & 1.65 & 1.35   \\
			8  & 9 & 2.8 & 2.2 & 1.8   \\		
			9  & 10 & 9 & 3 & 2.7   \\
			10  & 4 & 4 & 2 & 0   \\
			11  & 4 & 4 & 2.8 & 0    \\ 
			12  & 6 & 6 & 2 & 0     \\	 
			13  & 3.2 & 2.56 & 0.9 & 0   \\		
			14  & 6 & 4 & 1.6 & 0   \\
			
			\noalign{\smallskip}\hline\hline
	\end{tabular} } 	
\end{table}

In the strong inter-sublattice coupling regime, the critical condition separating isolated and condensed skyrmion phases is given by
$\kappa_{eff}=1$. This criterion allows the construction of a phase diagram in the $\kappa_{1_{eff}}\kappa_{2_{eff}}$-plane by identifying
the corresponding critical boundary. Using Eq. (\ref{E6}), the phase boundary condition $\kappa_{eff}=1$ yields the phase boundary equation of
\begin{equation}
	\kappa_{2_{eff}}=\frac{1}{\left( 1-\frac{1}{\kappa_{1_{eff}}}\right)},
	\label{E7}
\end{equation}
which separates condensed skyrmions from isolated skyrmions in the $\kappa_{1_{eff}}\kappa_{2_{eff}}$-plane.
According to Eq. (\ref{E7}), when $\kappa_{1_{eff}}\rightarrow\infty$ ($\kappa_{2_{eff}}\rightarrow\infty$), one has
$\kappa_{2_{eff}}=1$ ($\kappa_{1_{eff}}=1$), while both $\kappa_{1_{eff}}$ and $\kappa_{2_{eff}}$ must take
positive values. For the isolated skyrmion phase, $\kappa_{eff}<1$, which leads to $(1/\kappa_{1_{eff}}+1/\kappa_{2_{eff}})>1$ and
for the condensed skyrmion phase, $\kappa_{eff}>1$, which leads to $(1/\kappa_{1_{eff}}+1/\kappa_{2_{eff}})<1$. These inequalities
define the regions of isolated and condensed skyrmions in the $\kappa_{1_{eff}}\kappa_{2_{eff}}$-plane and directly determine
the phase boundary of the phase diagram. 

Numerically, we can either fix $\kappa_{1_{eff}}$ and vary $\kappa_{2_{eff}}$ by changing the $K_{u_2}$ or fix $\kappa_{2_{eff}}$ and
vary $\kappa_{1_{eff}}$ by changing $K_{u_1}$. It is straightforward to use MuMax3 to simulate skyrmions for a fixed set of model parameters,
and to use exactly the same methods as discussed above to find critical $\kappa_{1_{eff}}$ and $\kappa_{2_{eff}}$ that separate
isolated skyrmions from the condensed skyrmions. We use samples of size $800\,\mathrm{nm}\times800\,\mathrm{nm}\times d$, where $d = 2,
3\,\mathrm{nm}$ correspond to $(t_1, t_2) = (1, 1)\,\mathrm{nm}$ and $(1, 2)\,\mathrm{nm}$, respectively with the same
initial configurations used above to numerically determine the critical $\kappa_{eff}$. In our simulations, we keep $M_{s_1}= 0.58\, 
\mathrm{MA\,m^{-1}}$, $M_{s_2}= 0.7\, \mathrm{MA\,m^{-1}}$, $A_\ell$ and $D_\ell$ as in Set 9 in the Tab. \ref{T1},
and vary both $K_{u_1}$ and $K_{u_2}$ to change $\kappa_{1_{eff}}$ and $\kappa_{2_{eff}}$ values. The strong-coupling condition is
ensured by fixing $\zeta_{eff}=0.1$. Figure \ref{F2} displays (meta)stable skyrmions for various values of ($\kappa_{1_{eff}}, \kappa_{2_{eff}}$)
specified in the caption for the case of $t_1=t_2$.
Skyrmions in Fig. \ref{F2}(a-d) for sublattice 1, Fig. \ref{F2}(e-h) for sublattice 2 are circular and isolated while those in 
Fig. \ref{F2}(i-l) for sublattice 1, Fig. \ref{F2}(m-p) for sublattice 2 are stripe skyrmions. 
The phase boundary that separates isolated skyrmions from condensed
skyrmions passes through the middle of these pairs of points in $\kappa_{1_{eff}}\kappa_{2_{eff}}$-plane.
The time evolution of energy $E$ (blue curves in arcsinh scale) and net skyrmion
number $Q$ (red lines) shown in Fig. \ref{F2}(q-t) for above selected pairs of
($\kappa_{1_{eff}}, \kappa_{2_{eff}}$) supports the identification of the phase boundary.
\begin{figure}[h] 
	\centering
	\includegraphics[width=8cm]{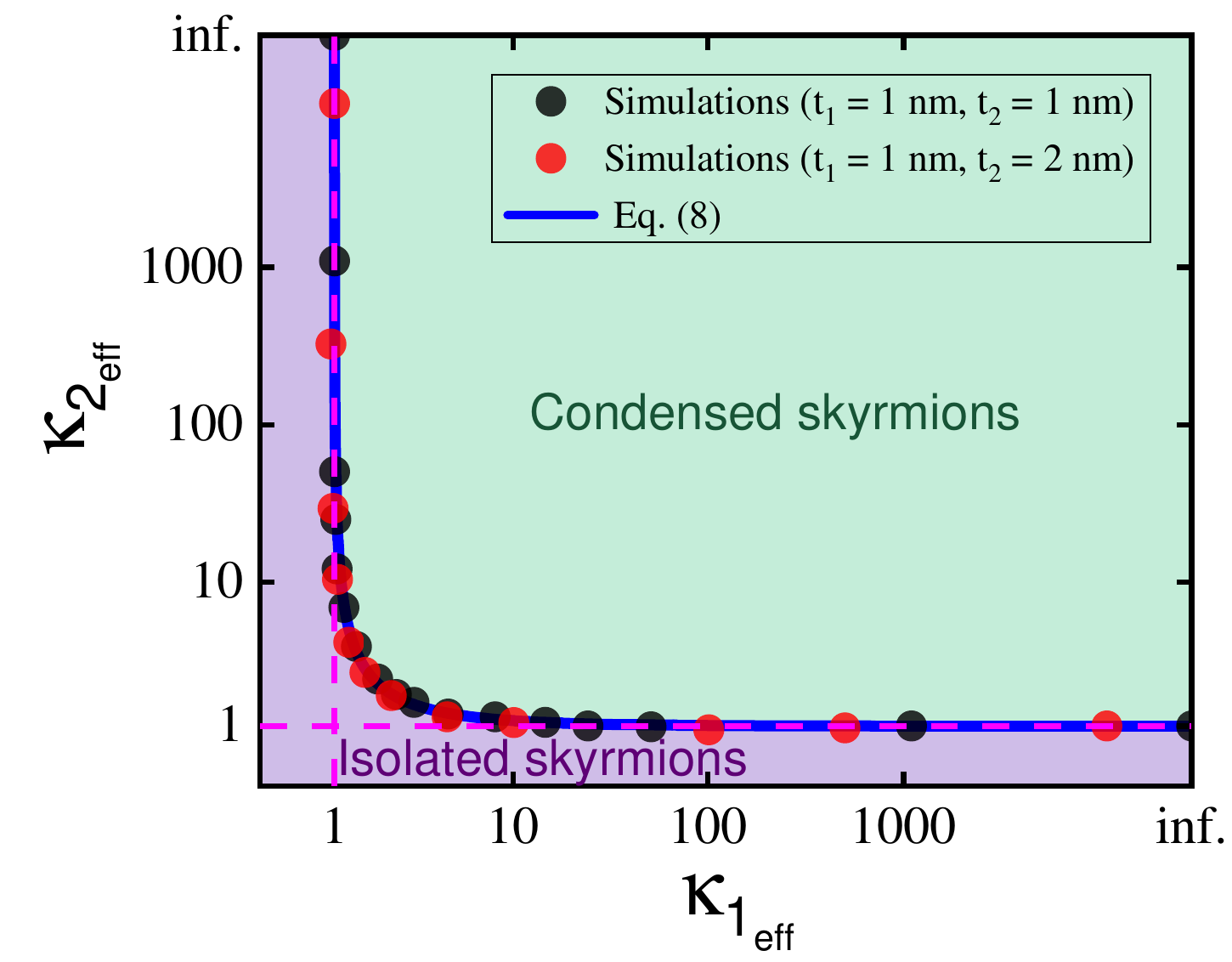} 
	\caption{Phase diagram of isolated and condensed skyrmions in the $\kappa_{1_{eff}}\kappa_{2_{eff}}$-plane in an arcsinh scale
		under strong coupling for different cases of $t_1/t_2$. The blue curve is the analytical results of
		the Eq. (\ref{E7}), and it perfectly align
		with symbols from simulations. The green regions is the phase of condensed skyrmions, the purple region is the phase for
		the isolated skyrmions. Phase boundary passes through $(1, \infty)$ and $(\infty, 1)$.}
	\label{F3}
\end{figure}

Based on the extensive numerical simulations, the phase diagram of
isolated skyrmions and condensed skyrmions in $\kappa_{1_{eff}}\kappa_{2_{eff}}$-plane is presented in
Fig. \ref{F3} for the cases of different $t_1/t_2$ ratios. $\kappa_{1_{eff}}$ and $\kappa_{2_{eff}}$ are plotted in arcsinh scale.
The curve between two pink
dashed lines is the phase boundary which separates isolated skyrmions from condensed skyrmions. The green region corresponds to
the condensed skyrmion phase, while the purple region is the
isolated skyrmion phase. The blue curve is the analytical results of Eq. (\ref{E7}). The black and red
symbols correspond to the cases $(t_1, t_2) = (1, 1)\,\mathrm{nm}$ and $(1, 2)\,\mathrm{nm}$, respectively
and align with blue curve perfectly, demonstrating the excellent
agreement between our simulations and the analytical results, which again validates the effective description and confirms
that the transition between isolated and condensed phases is governed by the single parameter $\kappa_{eff}=1$ in the 
strong-coupling regime. As expected,
($\kappa_{1_{eff}}=1, \kappa_{2_{eff}}=\infty$), ($\kappa_{1_{eff}}=\infty, \kappa_{2_{eff}}=1$) are clearly on the phase boundary,
in consistent with the Eq. (\ref{E7}). Furthermore, our simulations confirm that even when the thickness ratio $t_1/t_2$ is varied, the 
transition between isolated and condensed skyrmion phases consistently occurs at $\kappa_{eff} = 1$. This indicates that thickness does not act as an independent
control parameter for the phase transition itself, but rather modifies the effective material parameters entering $\kappa_{eff}$. Consequently,
changes in thickness influence quantities such as the characteristic length scale $L_{eff}=(4A_{eff})/(\pi D_{eff})$, and hence the size and internal
structure of skyrmions, without altering the critical condition for phase stability. This behavior reflects the effective-medium nature of the strongly
coupled FiM system, where the FiM skyrmion texture is fully described by the effective parameters, and the phase boundary is determined solely
by their dimensionless combination $\kappa_{\ell_{eff}}$. Importantly, as discussed in Sec. \ref{S3A*}, the coupling term $\zeta_{eff}$ does not
affect the skyrmion phase stability and does not change the phase boundary at $\kappa_{eff}=1$. To the best of our knowledge,
this case was not carefully studied before. Additionally, it should be noted
that besides the two phases shown in the phase diagram in Fig. \ref{F3} (isolated, and condensed skyrmion phases), there should be an
additional phase of the FiM state. This FiM state emerges
under strong magnetic anisotropy, where
isolated skyrmions shrink and ultimately transition into the FiM state.
Consequently, the FiM phase should occupy a narrow region near the $\kappa_{1_{eff}}$ and
$\kappa_{2_{eff}}$ axes in Fig. \ref{F3}. This suggests a need for further investigation, which has not
been examined in our current study.

\subsection{\texorpdfstring{Weak coupling regime and breakdown of locking}
		{Weak coupling regime and breakdown of locking}}
\label{S3B}
\begin{figure*}[t] 
	\centering
	\includegraphics[width=17.8cm]{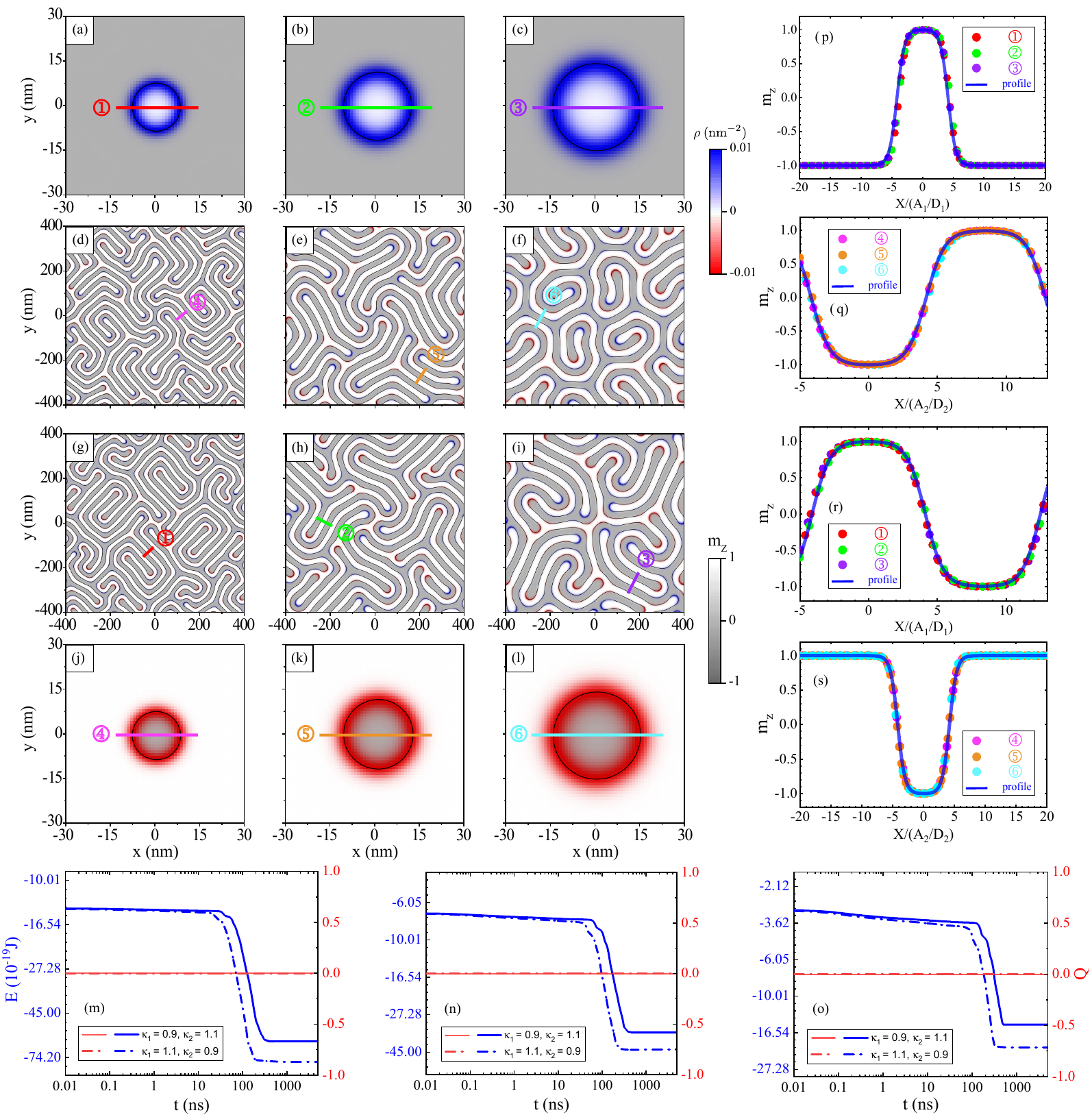} 
	\caption{(Meta)stable skyrmions for weak coupling regime. Skyrmions for (\textbf{a-c}) $\kappa_1= 0.9$ for sublattice 1,
		(\textbf{d-f}) $\kappa_2= 1.1$ for sublattice 2 and (\textbf{g-i}) $\kappa_1= 1.1$ for sublattice 1, (\textbf{j-l}) $\kappa_2= 0.9$ for sublattice 2 in a sample of
		$800\,\mathrm{nm}\times800\,\mathrm{nm}\times2\,
		\mathrm{nm}$ with saturation magnetization $M_{s_1}= 0.58\, \mathrm{MA\,m^{-1}}$ and $M_{s_2}= 0.7\, \mathrm{MA\,m^{-1}}$, and 
		exchange stiffness $A_\ell$ and DMI strength $D_\ell$ in Set 1 (\textbf
		{a, d, g, j}), Set 2 (\textbf{b, e, h, k}), and Set 3 (\textbf{c, f, i, l}) listed in Tab. \ref{T1}. 
		(\textbf{a-c}), (\textbf{j-l}) Isolated skyrmions. (\textbf{d-i}) Stripe skyrmions.  The color bar denotes skyrmion 
		charge density $\rho$ for each sublattice $\ell=1, 2$, and the gray-scale encodes $m_z$.
		(\textbf{m-o}) The net skyrmion number $Q$ (red lines) and energy $E$ (in arcsinh scale, blue curves) 
		vs. time $t$ (in logarithmic scale) for $\kappa_1= 0.9, \kappa_2=1.1$ (the solid lines) and $\kappa_1= 1.1, \kappa_2=0.9$ (the dashed-dot lines) 
		with $A_\ell$ and $D_\ell$ in Set 1 (\textbf{m}),  Set 2 (\textbf{n}), and Set 3 (\textbf{o}). 
		(\textbf{p-s}) $m_z$-distribution along the red (the red symbols), green (the green symbols), purple (the purple symbols), 
		pink (the pink symbols), orange (the orange symbols), and blue (the blue symbols) 
		lines in (\textbf{a-c}), (\textbf{d-f}), (\textbf{g-i}), and (\textbf{j-l}), respectively. 
		The solid blue lines are $\Theta(r)=2\arctan[\sinh(r/w)/\sinh(R/w)]$ with $R=4.15$ and $w=1.23$ 
		(\textbf{p}), $R=4.12$ and $w=1.2$ (\textbf{s}), and $\Theta(x)=2\arctan[\sinh(|x|/w)/\sinh(L/2w)]$ with $L=8.36$ and $w=1.36$ 
		(\textbf{q}), $L=8.24$ and $w=1.31$ (\textbf{r}) where $\Theta$ is 
		the polar angle of magnetization, $R$ and $L$ are skyrmion size, and $w$
		is the skyrmion wall thickness. $r = 0$ and $x = 0$ denote skyrmion centers. The $x$-axis is in the units of $A_\ell/D_\ell$ for each sublattice.}
	\label{F00}
\end{figure*}
In the weak inter-sublattice coupling limit, the AFM exchange is insufficient to rigidly lock the two sublattices,
and the magnetizations $\vec{m}_1$ and $\vec{m}_2$ relax largely independently. Under this limit, the energy of each sublattice is
evaluated independently for its own spin texture while the inter-sublattice coupling provides only a weak correction. This regime is conveniently characterized by the
small parameter $\varepsilon \equiv 1/\zeta_{eff} \ll 1$, which controls the strength of inter-sublattice locking. When
$\zeta_{eff} \gg 1$, the coupling term proportional to $\varepsilon$ enters the Eq. (\ref{E3}) as a weak perturbation and vanishes
to leading order. Consequently, each sublattice independently satisfies its own metastability condition,
$\vec{m}_\ell\times\{ a_\ell\nabla^2\vec{m}_\ell+\frac{4}{\pi}d_\ell[(\nabla\cdot\vec{m}_\ell)\hat{z}-\nabla 
	m_{z_\ell}]+\frac{1}{\kappa_{\ell_{eff}}}m_{z_\ell}\hat{z}\}=0$.
	Rearranging this, one can have,
$\vec{m}_\ell\times (\frac{d_\ell^2}{a_\ell}) \{( \frac{a_\ell}{d_\ell})^2  \nabla^2\vec{m}_\ell+\frac{4}{\pi}( \frac{a_\ell}{d_\ell}) [(\nabla\cdot\vec{m}_\ell)\hat{z}-\nabla 
	m_{z_\ell}]+(\frac{a_\ell}{d_\ell^2}) \frac{1}{\kappa_{\ell_{eff}}}m_{z_\ell}\hat{z}\}=0$.
By rescaling the spatial coordinates according to $x\to (L_{eff}/L_\ell)x$ and $y\to (L_{eff}/L_\ell)y$, this equation reduces to the standard dimensionless form,
\begin{equation}
	\vec{m}_\ell\times\left\lbrace \nabla^2\vec{m}_\ell+\frac{4}{\pi}[(\nabla\cdot\vec{m}_\ell)\hat{z}-\nabla 
	m_{z_\ell}]+\frac{1}{\kappa_{\ell}}m_{z_l}\hat{z}\right\rbrace=0.
	\label{Eweak}
\end{equation}
Here $\kappa_\ell = (\pi^2 D_\ell^2)/(16A_\ell K_\ell)$ and $L_\ell=(4A_\ell)/(\pi D_\ell)$, $x$ and $y$ are measured in the units of $L_{\ell}$ for each sublattice $\ell=1,2$.
According to the Eq. (\ref{Eweak}), in the weak-coupling limit,
skyrmion (meta)stability and phase boundaries are governed solely by
the intrinsic control parameter $\kappa_\ell$ of each sublattice, and the critical condition $\kappa_\ell=1$ separating the isolated and condensed phases, as established
for chiral FM films \cite{R10, R11, R12, R13}, while inter-sublattice coupling modifies the skyrmion texture without affecting the phase stability as
discussed in Sec. \ref{S3A*}.

To verify the critical value of $\kappa_{\ell}=1$, we use a sample of $800\,\mathrm{nm}\times800\,\mathrm{nm}\times2\,\mathrm{nm}$
with the same initial configuration as in Sec. \ref{S3A}. The initial
skyrmion configurations relax into the (meta)stable states for a given $K_{u_1}, K_{u_2}$, $M_{s_1}= 0.58\,
\mathrm{MA\,m^{-1}}$, $M_{s_2}= 0.7\, \mathrm{MA\,m^{-1}}$, and $A_\ell$, 
$D_\ell$ as specified in Set \# of Tab. \ref{T1}. A fixed $\zeta_{eff}= 1000$ is chosen to ensure weak inter-sublattice
coupling by choosing the corresponding $J$,
and the data are collected by fixing $\kappa_{\ell} = 0.9$ and $\kappa_{\ell}= 1.1$, which are below and above the critical
value separating isolated and condensed skyrmion phases. The resulting (meta)stable
spin structures obtained from the simulations are shown in Fig. \ref{F00}(a, d, g, j) for Set 1, Fig. \ref{F00}(b, e, h, k) for Set 2,
and Fig. \ref{F00}(c, f, i, l) for Set 3, where Fig. \ref{F00}(a-c), and \ref{F00}(g-i) show the (meta)stable 
spin structures of sublattice 1, while Fig. \ref{F00}(d-f), and \ref{F00}(j-l) show the spin structures of
sublattice 2.
For $\kappa_{1}= 0.9$ ($\kappa_{2}=0.9$), Fig. \ref{F00}(a-c) (Fig. \ref{F00}(j-l)) show isolated circular skyrmions,
whereas for $\kappa_{2} = 1.1$ ($\kappa_{1}=1.1$), Fig. \ref{F00}(d-f) (Fig. \ref{F00}(g-i))
show stripe skyrmions. These results follow directly the critical value $\kappa_{\ell} = 1$, which separates
isolated skyrmions from condensed skyrmions in the absence of an external magnetic field.
The skyrmion charge density 
$\rho$ is encoded by the colors specified in the color bar
for each sublattice. For a specific $\kappa_{\ell}$, the size of
the (meta)stable skyrmions varies with $A_{\ell}$ and $D_{\ell}$, or with the length scale $L_{\ell}=(4A_{\ell})/(\pi D_{\ell})$.
To further validate this length scale, the $m_z$-distribution is examined along the red (red symbols), green (green symbols),
and purple (purple symbols) lines in Fig. \ref{F00}(a–c) for sublattice 1 and along the pink (pink symbols), orange (orange symbols),
blue (blue symbols)  lines in Fig. \ref{F00}(d–f) for sublattice 2, are plotted as shown in Fig. \ref{F00}(p), \ref{F00}(q), respectively.
Similarly, the $m_z$-distribution along the same sets of lines in Fig. \ref{F00}(g–i) for sublattice 1 and Fig. \ref{F00}(j–l) for
sublattice 2 are plotted as shown in Fig. \ref{F00}(r), \ref{F00}(s), respectively.
The symbols are simulation data obtained from
MuMax3 and the $x$-axis is scaled by $A_{l}/D_{l}$. The spin profiles are well described by $\Theta(r)=2\arctan[\sinh(r/w)/\sinh(R/w)]$
for isolated circular
skyrmions \cite{XS2018} and $\Theta(x)=2\arctan[\sinh(|x|/w)/\sinh(L/2w)]$ for stripe skyrmions \cite{R10} as shown by the blue curves in
Fig. \ref{F00}(p), \ref{F00}(q) and \ref{F00}(r), \ref{F00}(s), respectively. The excellent agreement between simulations (symbols)
and theoretical profile (curves) confirms both the length scale and the universality of the skyrmion spin
profiles\cite{R5, R10}. The (meta)stability and skyrmion nature of
these textures are further confirmed by the
relaxation dynamics shown in Fig. \ref{F00}(m-o), where the net skyrmion charge $Q$ (the red lines and the right $y$-axis) and total
energy $E$ (the blue curves and the left $y$-axis) are plotted as functions of time for $\kappa_{1} = 0.9, \kappa_{2}=1.1$
(the solid curves) and $\kappa_{1} = 1.1, \kappa_{2}=0.9$ (the dashed-dot curves). In all cases, the net skyrmion number remains constant at $Q = 0$ for $t>10$ ps.
\begin{figure}[h] 
	\centering
	\includegraphics[width=8cm]{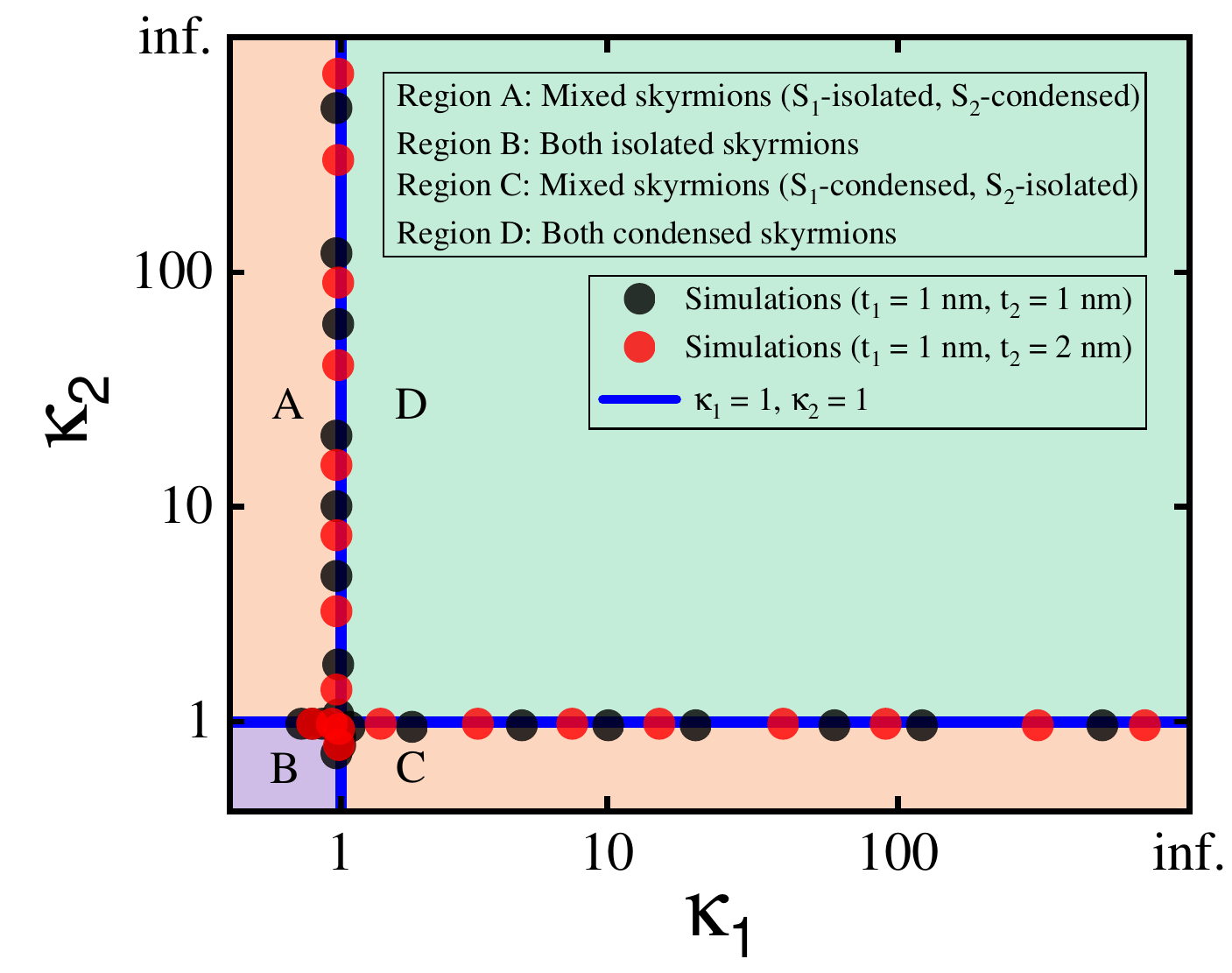} 
	\caption{Phase diagram of isolated and condensed skyrmions in the $\kappa_{1}\kappa_{2}$-plane in an arcsinh scale
			under weak coupling for different cases of $t_1/t_2$. The blue curves are the analytical results of
			the $\kappa_{1}=1, \kappa_{2}=1$, and it perfectly align with symbols from simulations. Regions A, B, C, and D correspond
			to mixed skyrmions ($\mathrm{S}_1$-isolated, $\mathrm{S}_2$-condensed), both isolated skyrmions, mixed skyrmions ($\mathrm{S}_1$-condensed, $\mathrm{S}_2$-isolated),
			and both condensed skyrmions, respectively.}
	\label{F9}
\end{figure}

In the weak inter-sublattice coupling regime, the critical condition separating isolated and condensed skyrmion phases is given by $\kappa_\ell = 1$. This allows the
construction of a phase diagram in the $\kappa_{1}\kappa_{2}$-plane, where the lines $\kappa_{1} = 1$ and $\kappa_{2} = 1$ separate condensed ($\kappa_{\ell} > 1$) and isolated
($\kappa_{\ell} < 1$) skyrmions in sublattices 1 and 2, respectively. Using the same approach as in the strong-coupling regime (Sec. \ref{S3A}), we construct the phase diagram by
fixing $A_\ell$ and $D_\ell$ from Set 6 in Tab. \ref{T1} and varying $K_{u_1}$ and $K_{u_2}$ to change $\kappa_{1}$ and $\kappa_{2}$, while ensuring the weak-coupling condition by
setting $\zeta_{eff} = 1000$. The resulting phase diagram is shown in Fig. \ref{F9} for different $t_1/t_2$ ratios, with $\kappa_{1}$ and $\kappa_{2}$ plotted in arcsinh scale.
Region A corresponds to mixed skyrmion states, where sublattice 1 hosts isolated skyrmions and sublattice 2 hosts condensed skyrmions. Region B corresponds to isolated skyrmion
states in both sublattices, where each sublattice independently supports isolated skyrmion configurations. Region C represents the opposite mixed state, where sublattice 1 hosts
condensed skyrmions and sublattice 2 hosts isolated skyrmions. Region D corresponds to condensed skyrmion states in both sublattices. The analytical boundaries $\kappa_{1} = 1$
and $\kappa_{2} = 1$ (blue lines) are in excellent agreement with simulation results for $(t_1, t_2) = (1, 1)$ nm (black) and $(1, 2)$ nm (red). These results confirm that the
phase boundary remains at $\kappa_{\ell} = 1$ even when the thickness ratio $t_1/t_2$ is varied. The FiM phase is expected to lie in a narrow region near the $\kappa_{1}$
and $\kappa_{2}$ axes in Fig. \ref{F9}, which is not explored in the present study.

\subsection{\texorpdfstring{Skyrmion stabilization in a DMI-free sublattice by inter-sublattice locking}
	{Skyrmion stabilization in a DMI-free sublattice by inter-sublattice locking}}
\label{S3C}
\begin{figure*}[t] 
	\centering
	\includegraphics[width=17.8cm]{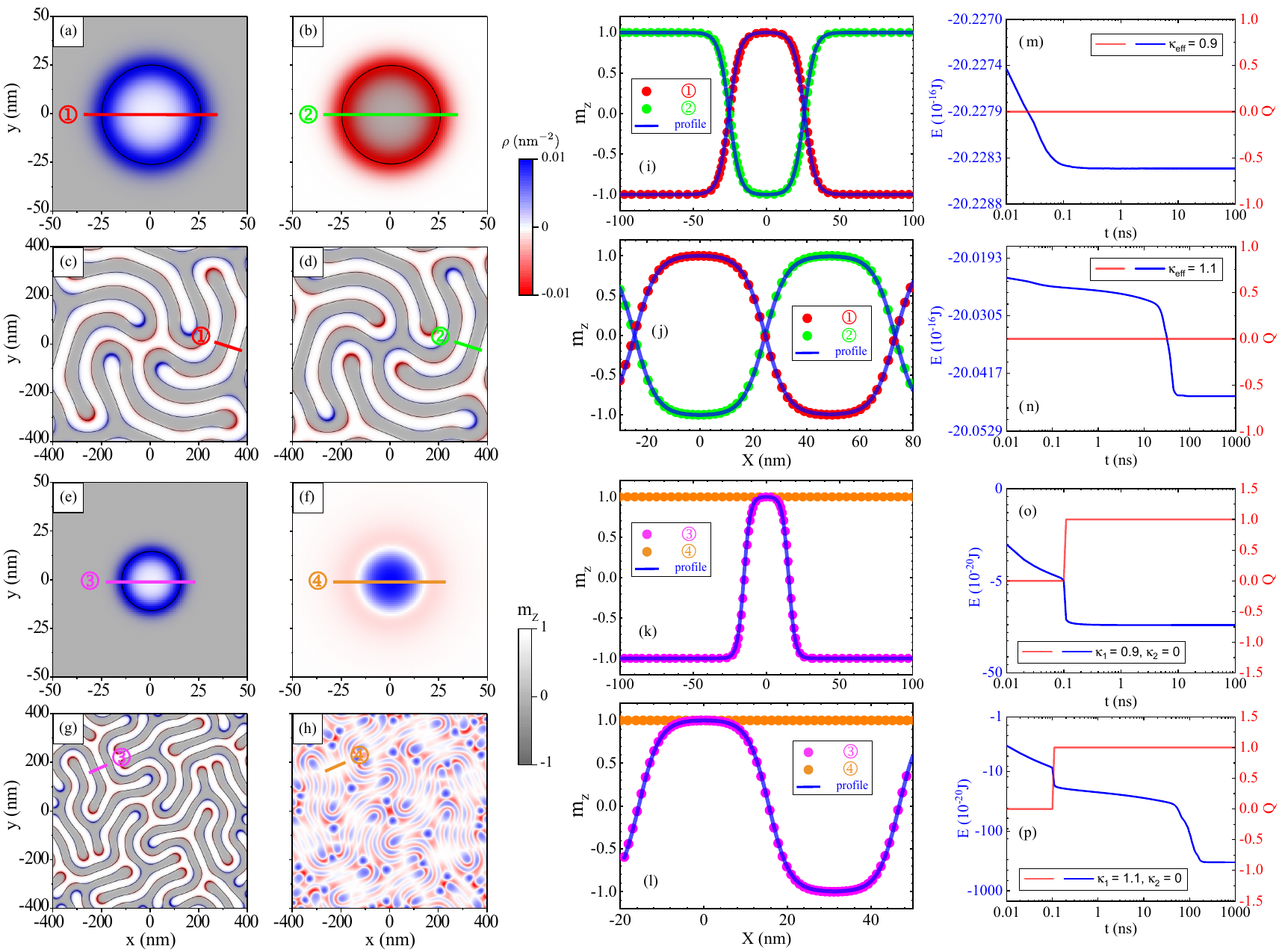} 
	\caption{(Meta)stable states in the finite DMI ($D_1 \neq 0$) and DMI-free ($D_2 = 0$) sublattices. (\textbf{a-d}) (Meta)stable states
		correspond to the strong inter-sublattice coupling
		regime, while (\textbf{e-h}) to the weak-coupling regime. (\textbf{a, c, e, g}) show sublattice 1 and
		(\textbf{b, d, f, h}) show
		sublattice 2 in a sample of $800\,\mathrm{nm}\times800\,\mathrm{nm}\times2\,
		\mathrm{nm}$ with saturation magnetization $M_{s_1}= 0.58\, \mathrm{MA\,m^{-1}}$ and $M_{s_2}= 0.7\, \mathrm{MA\,m^{-1}}$, and 
		exchange stiffness $A_\ell$ and DMI strength $D_\ell$ in Set 14 listed in Tab. \ref{T1} for $t_1=t_2$. In the strong-coupling regime both
		sublattices remain locked and form a FiM skyrmion: (\textbf{a, b}) isolated skyrmions ($\kappa_{eff}=0.9$) and
		(\textbf{c, d}) stripe skyrmions ($\kappa_{eff}=1.1$).
		In the weak-coupling regime, the locking breaks down and sublattice 1 supports
		isolated skyrmion ($\kappa_{1}=0.9$) (\textbf{e}) and stripe skyrmion ($\kappa_{1}=1.1$) (\textbf{g}), whereas sublattice 2
		(\textbf{f, h}) no longer hosts a skyrmion ($\kappa_{2}=0$)
		and instead relaxes toward a nearly uniform or modulated background state. The color bar denotes skyrmion charge density $\rho$
		for each sublattice $\ell=1, 2$,
		and the gray-scale encodes $m_z$.
		(\textbf{i-l}) $m_z$-distribution along the red (the red symbols), green (the green symbols) lines in (\textbf{a-d}), and pink
		(the pink symbols), orange (the orange symbols) lines in (\textbf{e-h})
		respectively. The solid blue lines are $\Theta(r)=2\arctan[\sinh(r/w)/\sinh(R/w)]$ with $R=(25.74, 25.54)\, \mathrm{nm}$ and
		$w=(7.28, 7.22)\, \mathrm{nm}$ (\textbf{i}), $R=15.33\, \mathrm{nm}$ and $w=4.37\, \mathrm{nm}$ (\textbf{k}), and
		$\Theta(x)=2\arctan[\sinh(|x|/w)/\sinh(L/2w)]$ with $L=(49.03, 49.03)\, \mathrm{nm}$ and $w=(8.23, 8.12)\, \mathrm{nm}$ 
		(\textbf{j}), $L=31.14\, \mathrm{nm}$ and $w=4.95\, \mathrm{nm}$ (\textbf{l}) where $\Theta$ is 
		the polar angle of magnetization, $R$ and $L$ are skyrmion size, and $w$
		is the skyrmion wall thickness. $r = 0$ and $x = 0$ denote skyrmion centers. (\textbf{m-p}) The net skyrmion number $Q$ (the red lines
		and the right $y$-axis) and energy $E$ in arcsinh 
		scale (the blue curves and the left $y$-axis) as a function of time (in logarithmic scale) for (\textbf{a, b}),
		(\textbf{c, d}), (\textbf{e, f}), and (\textbf{g, h}) respectively. Net skyrmion number is constant at $Q = 0$
		for strong-coupling regime and $Q = 1$ for weak-coupling regime.}
	\label{F7}
\end{figure*}
In conventional chiral magnets, the stabilization of skyrmions relies on the presence of DMI, which provides the necessary chiral
twisting to compete against exchange stiffness and anisotropy. In FiM systems, however, the presence of two antiferromagnetically
coupled sublattices introduces an additional stabilization mechanism that is absent in ferromagnets. Here, we demonstrate that
skyrmions can be stabilized even when one sublattice is DMI-free ($D_2 = 0$) while the other sublattice retains finite DMI
($D_1 \neq 0$), provided the inter-sublattice exchange coupling is sufficiently strong. In the strong-coupling regime, the two
sublattices remain rigidly locked and behave as a FiM skyrmion structure, as discussed in Sec. \ref{S3A}. In this limit,
the system can be described by effective parameters, including an effective DMI strength $D_{eff} = (D_1t_1 + D_2t_2)/d$. Consequently,
even when $D_2 = 0$, the system retains a finite $D_{eff} = D_1t_1/d$, which provides the necessary chiral twisting to
stabilize skyrmions in both sublattices. Thus, the DMI-free sublattice inherits chiral stabilization through strong inter-sublattice
exchange locking. As the AFM inter-sublattice coupling $J$ is reduced and the system enters the weak-coupling regime, this effective
description breaks down and the effective DMI picture is no longer valid because the two sublattices can relax independently as we
discussed in Sec. \ref{S3B}. In this regime, the DMI-free sublattice lacks intrinsic chiral interactions and therefore can no longer
sustain a skyrmion, leading to its collapse. This demonstrates that skyrmion stabilization in a DMI-free FiM sublattice is a
consequence of strong inter-sublattice locking and disappears once the coupling becomes insufficient to enforce a FiM skyrmion.
\begin{figure}[h] 
	\centering
	\includegraphics[width=8.4cm]{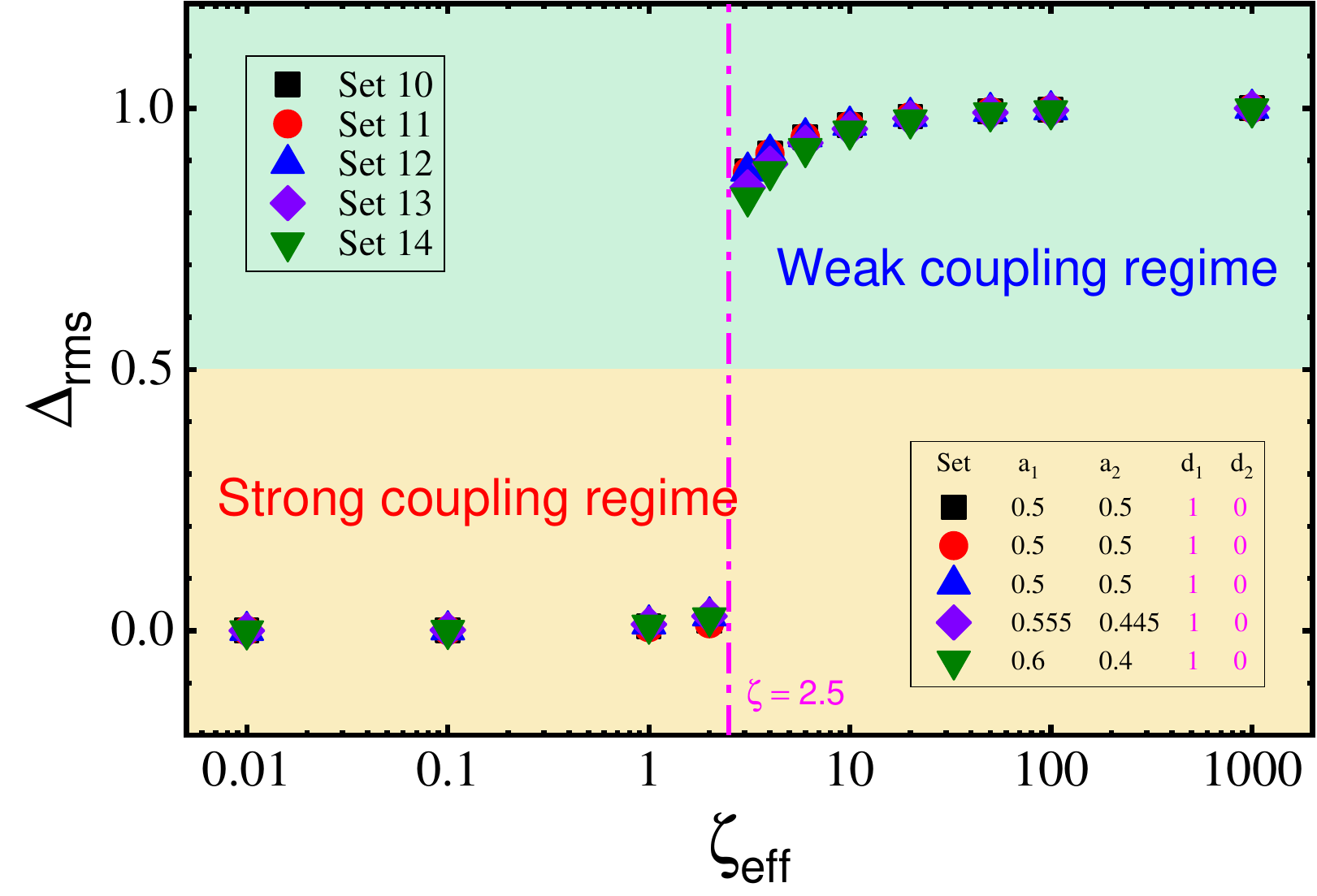} 
	\caption{Normalized order parameter, $\Delta_{rms}$ as a function of the effective inter-sublattice
		coupling strength $\zeta_{eff}$ (in logarithmic scale) in the DMI-free sublattice ($D_2 = 0$) for several representative
		material parameter sets $A_\ell, D_\ell$ in Set 10 (rectangles), Set 11 (circles), Set 12 (up-triangles), Set 13 (diamonds), Set 14
		(down-triangles) listed in Tab. \ref{T1} for $t_1 = t_2$ with fixed $\kappa_{1_{eff}} = 1.6$ and $\kappa_{2_{eff}} = 2.2$. A sharp
		crossover separates the strong-coupling regime ($\zeta_{eff}<2.5$), from
		the weak-coupling regime ($\zeta_{eff}>2.5$).}
	\label{F6}
\end{figure}

To verify the existence of skyrmion in the DMI-free limit of sublattice 2 ($D_2 = 0$), we use MuMax3 to numerically determine the
(meta)stable states in both strong and weak inter-sublattice coupling regimes in a sample size of
$800\,\mathrm{nm}\times800\,\mathrm{nm}\times2\,\mathrm{nm}$ with $M_{s_1}= 0.58\, \mathrm{MA\,m^{-1}}$, $M_{s_2}= 0.7\,
\mathrm{MA\,m^{-1}}$, and $A_\ell$ and $D_\ell$ in the Set 14 listed in Tab. \ref{T1} for the case of $t_1=t_2$. The initial configuration
is the same as that in Sec. \ref{S3A}.  The strong-coupling regime is ensured by setting $\zeta_{eff}=0.1$ and the weak-coupling regime by setting
$\zeta_{eff}=1000$ through the corresponding choice of $J$. In the strong-coupling case,
$\kappa_{eff}=0.9$ and $\kappa_{eff}=1.1$ are obtained by changing $K_{u_1}$ and $K_{u_2}$, while in the weak-coupling regime,
$K_{u_1}$ is changed to obtain $\kappa_1 = 0.9$, $\kappa_1 = 1.1$, and a finite $K_{u_2}$ is chosen such that $\kappa_2 = 0$ since
$D_2 = 0$,  allowing a systematic  comparison between the strong and weak coupling regimes. In the strong inter-sublattice coupling
regime, the resulting (meta)stable states are shown in Fig. \ref{F7}(a–d) where sublattice 1,
shown in Fig. \ref{F7}(a, c), and sublattice 2, shown in Fig. \ref{F7}(b, d), remain rigidly locked where isolated
skyrmion for $\kappa_{eff}=0.9$ in Fig. \ref{F7}(a, b) and stripe skyrmion for $\kappa_{eff}=1.1$ in Fig. \ref{F7}(c, d).
In contrast, when the inter-sublattice exchange coupling is reduced and the system enters the weak-coupling regime shown in Fig.
\ref{F7}(e–h), the locking between the two sublattices breaks down. Sublattice 1, shown in Fig. \ref{F7}(e, g), continues to host an
isolated skyrmion for $\kappa_{1}=0.9$ in Fig. \ref{F7}(e) and a stripe skyrmion for $\kappa_{1}=1.1$ in Fig. \ref{F7}(g), owing to
its finite DMI. However, sublattice 2, shown in Fig. \ref{F7}(f, h), corresponding to $\kappa_{2}=0$, lacks
intrinsic chiral interactions and relaxes toward nearly uniform or modulated background states, no longer hosting skyrmions.
The $m_z$-distribution of sublattices 1, 2 is examined along the red (red symbols), green (green symbols) lines in Fig. \ref{F7}(a, b),
Fig. \ref{F7}(c, d), and along the pink (pink symbols),  orange (orange symbols) lines in Fig. \ref{F7}(e, f), Fig. \ref{F7}(g, h)
and plotted as shown in Fig. \ref{F7}(i), \ref{F7}(j), \ref{F7}(k), and \ref{F7}(l), respectively. The symbols are simulation data
obtained from MuMax3 and the $x$-axis is in nm range. The spin profiles are well described by $\Theta(r)=2\arctan[\sinh(r/w)/\sinh(R/w)]$
for isolated circular skyrmions \cite{XS2018} and $\Theta(x)=2\arctan[\sinh(|x|/w)/\sinh(L/2w)]$ for stripe skyrmions \cite{R10} as
shown by the blue curves in Fig. \ref{F7}(i), \ref{F7}(j) and \ref{F7}(k), \ref{F7}(l), respectively. 
The isolated circular skyrmion sizes of sublattices 1 and 2, $R = (25.74, 25.54)$ nm
and the stripe skyrmion sizes $L = (49.03, 49.03)$ nm indicate that the skyrmions in both sublattices share the same characteristic
size under strong-coupling in the DMI-free limit. The (meta)stability and skyrmion nature of these textures are further confirmed by the
relaxation dynamics shown in Fig. \ref{F7}(m-p), where the net skyrmion charge $Q$ (the red line and the right $y$-axis) and total
energy $E$ (the blue curve and the left $y$-axis) are plotted as functions of time for $\kappa_{eff} = 0.9$, $\kappa_{eff} = 1.1$,
$\kappa_{1} = 0.9$, and $\kappa_{1} = 1.1$  (the solid curves) respectively. In all cases, the net skyrmion
number remains constant at $Q = 0$ under strong coupling and $Q=1$ for $t>0.1$ ns under weak coupling, while the total energy converges to
a stable value. The apparent spatial features in sublattice 2 under weak coupling do not correspond to a stabilized skyrmion. Although
weak localized contrast appears in the topological charge density, the $m_z$-distribution remains at 1 and does not exhibit polarity reversal,
indicating that no skyrmion core is formed and the magnetization does not
wrap the unit sphere.
The finite local charge density arises from weak exchange-induced canting inherited from the skyrmion in sublattice 1; however, these
contributions cancel upon integration, yielding zero net skyrmion number for sublattice 2. The dynamical evolution further confirms
this picture: while the net skyrmion number of the system change from $Q=0$ to $Q=1$ during relaxation for $t>0.1$ ns, the resulting
skyrmion is hosted solely by sublattice 1, whereas sublattice 2 relaxes to a nearly uniform
background with only weak, non-topological distortions. This results confirm that FiM skyrmions remain stable in the
strong-coupling regime but collapse in the DMI-free sublattice once the coupling becomes insufficient to enforce the collective behavior.

Furthermore, we plot the $\Delta_{rms}$ as a function of $\zeta_{eff}$ considering the case of $t_1=t_2$
using the same procedure described in Sec. \ref{S3A*}
for a fixed $\kappa_{1_{eff}}=1.6$, $\kappa_{2_{eff}}=2.2$,
as shown in Fig. \ref{F6}. Figure \ref{F6} demonstrates that the emergence of skyrmion in a DMI-free sublattice is governed
by the effective inter-sublattice
coupling strength $\zeta_{eff}$ across a broad range of material parameters ($A_\ell, D_\ell$ as specified in Set \# of Tab. \ref{T1}). 
When $D_2 = 0$, $d_\ell$ becomes identical for the entire parameter space and does not contribute to the shift of the crossover.
As a result, the critical crossover value is shifted by $a_\ell$ for a given $\kappa_{\ell_{eff}}$, which is known to induce only a very narrow variation. Consequently, across
the entire parameter space, the critical crossover narrows and converges to $\zeta_{eff} \approx 2.5$.
In the strong-coupling regime
($\zeta_{eff} < 2.5$), the two sublattices remain rigidly antiparallel and behave as a FiM skyrmion, resulting in
$\Delta_{rms} \approx 0$. In this regime, even though $D_2 = 0$, the effective locking between the sublattices enables stabilization
of a skyrmion texture in the DMI-free sublattice through inter-sublattice exchange. As $\zeta_{eff}$ increases beyond the critical
value $\zeta_{eff} \approx 2.5$, a sharp crossover occurs to a weak-coupling regime ($\zeta_{eff}>2.5$) characterized by finite
angular deviation between
the sublattice magnetizations, where the effective description breaks down and skyrmions in the DMI-free sublattice are no longer sustained.
Therefore, skyrmions can be stabilized in both sublattices even when one is a DMI-free sublattice under the condition $\zeta_{eff} < 2.5$.
It should be noted, as the critical crossover depends on the $\kappa_{\ell_{eff}}$
values of the two sublattices, $\zeta_{eff} \approx 2.5$ obtained here shift as $\kappa_{\ell_{eff}}$ varies.

\section{Discussion and Conclusion}
\label{S4}

We explore the phase diagram of isolated and condensed skyrmions in FiM films over a broad parameter space defined by the sublattice specific parameters
and the AFM inter-sublattice coupling in the absence of a magnetic field. Our analysis demonstrates that the crossover between strong and weak coupling regimes
is governed primarily by $\zeta_{eff}$, and $\kappa_{\ell_{eff}}$ primarily governs the
skyrmion phase stability while determining where the crossover occurs together with $a_\ell$ and $d_\ell$. Based on the existing understanding that the parameter $\kappa$
determines the (meta)stable spin textures,
with the natural length scale $L$, in FM films, we construct the phase diagram in the $\kappa_{1_{eff}}$$\kappa_{2_{eff}}$-plane for the strong-coupling
regime using the effective $\kappa$-based description. This effective description, however, breaks down in the weak-coupling regime as the $J$ decreases, and the
(meta)stable structures of each sublattice are then determined by their intrinsic material parameters. In the strong inter-sublattice coupling regime, the antiparallel
locking reduces the system to a FiM skyrmion governed by effective parameters. In this limit, the energy functional simplifies to a form
equivalent to that of a chiral ferromagnet with renormalized coefficients lead to a single effective equation identical in structure to that of a
chiral ferromagnet, but with parameters emerging from thickness-weighted averaging, $\kappa_{eff}$ that separate isolated ($\kappa_{eff}<1$) and
condensed ($\kappa_{eff}>1$) skyrmion phases. The phase boundary between the isolated and condensed skyrmions pass through the points $(1, \infty)$ and
$(\infty, 1)$ in the $\kappa_{1_{eff}}\kappa_{2_{eff}}$-plane supports this effective description, with $\kappa_ {eff} = 1$ solely controlling the transition
between isolated and condensed skyrmion phases. In addition, the agreement between the phase boundaries obtained for different thickness combinations further
highlights the physical relevance of this effective description. Although the two sublattices can have unequal thicknesses, which modifies their relative
contributions to the effective parameters through thickness weighting, the transition between isolated and condensed skyrmion phases remains governed
solely by $\kappa_{eff} = 1$. Also, for a given $\kappa_{eff}$, the size of the (meta)stable skyrmions is determined by the effective length scale $L_{eff}$.
Importantly, this effective description explains why skyrmion phases in ferrimagnets may resemble those in ferromagnets under strong locking, while still
retaining sublattice asymmetry at the microscopic level. In the weak-coupling regime, the effective $\kappa$-based description no longer applies.
The two sublattices are governed independently by their intrinsic control parameters $\kappa_{\ell}=1$ separating isolated ($\kappa_{\ell}<1$) and condensed
($\kappa_{\ell}>1$) skyrmion phases, and the stability of skyrmions must be determined separately for each sublattice. This independent behavior allows
a natural representation of the phase stability in terms of a phase diagram in the $\kappa_{1}\kappa_{2}$-plane. The lines $\kappa_{1} = 1$ and $\kappa_{2} = 1$ separate
the plane into four regions that reflect the skyrmion morphology in each sublattice. Together, these lines separate the parameter space into distinct regimes corresponding
to different combinations of skyrmion phases in the two sublattices, where both sublattices host isolated skyrmions, both host condensed
skyrmions, or mixed skyrmions where one sublattice supports isolated skyrmions while the other supports condensed skyrmions or vice versa. Also, for a given $\kappa_{\ell}$,
the length scale of each sublattice $L_{\ell}$ determines the size of its (meta)stable skyrmions. Our findings provide a clear method to experimentally
distinguish isolated and condensed skyrmions in FiM systems. This distinction is important
for practical applications, as identifying the underlying skyrmion phase can guide the design and operation of spintronic devices. Furthermore, these findings
provide a unified framework linking inter-sublattice exchange to skyrmion phase stability and offer practical design principles for coupling-engineered
spintronic devices based on FiM systems. As a representative example, while it is generally understood that skyrmions in chiral magnetic films
require DMI for stabilization, we find a striking result
that skyrmions can remain stabilized in one sublattice even when its intrinsic DMI is zero, provided that the other sublattice has a finite DMI in the strong-coupling
regime. In this regime, the system retains a finite effective DMI, so that even when one sublattice has no DMI and the other has finite DMI, the system
remains chiral at the effective level. The DMI-free sublattice does not generate chirality on its own, but instead follows the twisting through rigid
inter-sublattice exchange locking. As the coupling weakens, this effective description breaks down, and the DMI-free sublattice relaxes toward a nearly
uniform configuration, while skyrmions remain only in the sublattice with finite DMI. Importantly, the crossover between the strong- and weak-coupling regimes
becomes narrower in this case for given $\kappa_{1_{eff}}$ and $\kappa_{2_{eff}}$, since the relative DMI contributions $d_\ell$ remain fixed
(i.e., $d_1=1, d_2=0$ or vice versa) across the parameter space, with one sublattice being DMI-free. As a result, the tendency for chiral twisting remains unchanged,
which suppresses the shifting of the crossover and leads to a more clearly defined crossover location. In addition, layer thickness provides a direct and
experimentally accessible control parameter in multilayer FiM systems, as it can be precisely
tuned during fabrication to systematically modify the effective magnetic interactions. Varying the thickness of individual magnetic layers leads to a thickness-weighted
effective description, where the relative contributions of the two sublattices to the total energy are adjusted without changing their intrinsic material parameters.
As a result, effective material parameters can be continuously tuned through thickness alone, enabling controlled access to different skyrmion regimes.
The work of Luo et al. \cite{R_Luo2024} shows that the skyrmion size remains unchanged with layer thickness when the two sublattices
have identical material parameters. This observation is consistent with our results, as it represents a specific case of the general effective description
developed here. In contrast, we further show that when the material parameters of the two sublattices differ, the skyrmion size varies systematically with
layer thickness, demonstrating the broader applicability of our framework. This offers a practical route through thickness engineering to control the skyrmion
size and tune the system across the crossover between strong- and weak-coupling regimes identified in our work.

Our work is directly relevant to rare-earth–transition-metal ferrimagnets and synthetic AFM multilayers, where two magnetic sublattices
are antiferromagnetically coupled but possess different intrinsic material parameters. In FiM alloys, the exchange stiffness,
anisotropy, and DMI strength of the rare-earth and transition-metal sublattices are generally unequal \cite {R_Woo2018}. Since DMI originates from
heavy-metal interfaces \cite{R13}, its strength can couple differently to the two magnetic layers. Therefore, the sublattice dependent
parameters $A_\ell$, $D_\ell$, and $K_{u_\ell}$ in our model correspond naturally to experimentally controllable quantities through composition and interface engineering. 
In these FiM systems, the inter-sublattice exchange coupling depends sensitively on alloy composition and temperature,
particularly near magnetic or angular momentum compensation points \cite{R_Woo2018, R18}. This provides a practical way to vary the relative
strength of inter-sublattice locking without changing the overall film structure. As a result, an experimental system can move between regimes
where the two sublattices form a rigid FiM skyrmion and regimes where they relax more independently. The separation between strong and
weak coupling identified in our work therefore corresponds to experimentally accessible conditions.
A similar situation arises in synthetic antiferromagnets, where two FM layers are coupled through a spacer layer. The interlayer exchange
can be tuned continuously by varying spacer thickness, while DMI may differ between the two layers depending on their proximity to heavy metals \cite {R_Legrand2020}. 
In this case, spacer thickness acts as a direct control parameter for modifying the effective coupling between layers.
Our framework provides a systematic way to determine when such a bilayer can be described as a FiM skyrmion and when a full
two-layer description is necessary.

Our phase diagrams in the $\kappa_{1_{eff}}\kappa_{2_{eff}}$-plane in the strong-coupling regime and $\kappa_{1}\kappa_{2}$-plane in the weak-coupling regime
provide a complete and general description for
FiM films at zero temperature. This approach offers a clear and unified framework to understand and organize both experimental observations and
numerical results. With these phase diagrams, one can directly determine the phase behavior of a new material without performing extensive simulations across
a large parameter space. This greatly enhances the ability to explore the (meta)stability of skyrmion phases, since each point in the
$\kappa_{1_{eff}}\kappa_{2_{eff}}$-plane and $\kappa_{1}\kappa_{2}$-plane uniquely determine the corresponding (meta)stable spin structure. Therefore, our
phase diagrams provide a powerful and practical tool for guiding both theoretical studies and experimental design in FiM skyrmion systems
	
\begin{acknowledgements}
This work is supported by the NSFC Grant (No. 12374122), the Guangdong Provincial Quantum Strategy 
Special Project, and the University Development Fund of the Chinese University of Hong Kong (Shenzhen).  
M. V. Wijethunga acknowledges the support from the Hong Kong PhD Fellowship.
\end{acknowledgements}

\end{document}